%% file: Phase-field_damage.tex
\documentclass[final,3p,number,sort&compress]{elsarticle}

\pdfoutput=1

\usepackage[english]{babel}
\usepackage[utf8]{inputenc}
\usepackage[T1]{fontenc}
\usepackage{charter}
\usepackage{setspace}

\usepackage{graphicx,calc}
\usepackage[export]{adjustbox}
\usepackage{color,xcolor}
\definecolor{grau}{HTML}{6F6F6F}

\usepackage{tikz}
\tikzstyle{line}=[draw, thick, -stealth]

\usepackage[font=footnotesize]{caption}
\usepackage[font=footnotesize]{subcaption}
\usepackage{float}
\usepackage{listliketab}

\usepackage{tabularx}
\usepackage{array}
\usepackage{multirow}
\usepackage{booktabs}

\usepackage{amsmath}
\usepackage{amssymb}
\usepackage{amsfonts}
\usepackage{amsxtra}
\usepackage{mathtools}
\usepackage{empheq}
\usepackage{stmaryrd}
\usepackage{icomma}
\usepackage{relsize}

\usepackage[colorlinks=false,hidelinks]{hyperref}

\renewcommand\fbox{\fcolorbox{white!0}{white!0}} 
\newcolumntype{Y}{>{\raggedright\arraybackslash}X}
\newcolumntype{L}[1]{>{\raggedright\let\newline\\\arraybackslash\hspace{0pt}}p{#1}}
\newcolumntype{P}[1]{>{\centering\arraybackslash}m{#1}}

\newcommand{\grad}{\boldsymbol{\nabla}}

\input{defs}
\journal{arXiv}

\title{Numerical investigation of fracture behaviour of polyurethane adhesives under the influence of moisture}

\author{Siva~Pavan~Josyula} \ead{siva.josyula@uni-saarland.de}
\author{Stefan~Diebels} \ead{s.diebels@mx.uni-saarland.de}

\address{Chair of Applied Mechanics, Saarland University, Saarbr\"ucken, Germany}
\begin{document}
\begin{frontmatter}
	\begin{abstract}
		Polyurethane adhesives are extensively used in manufacturing of the lightweight components. These adhesives are hygroscopic in nature causing the material to degrade their mechanical properties due to ageing under environmental conditions. Polyurethane adhesive shows non-linear rate-dependent behaviour due to viscoelastic properties and assumes nearly incompressible deformation. Finite-strain viscoelastic model is formulated based on the continuum rheological model by combining spring and Maxwell elements in parallel. Nearly incompressible property is considered by decomposing the mechanical free energy into volumetric and isochoric parts. This work focuses on the investigation of the influence of moisture on the fracture behaviour. Phase-field damage model proved to be an efficient method to investigate the fracture behaviour. Therefore, the finite-strain viscoelastic material model is coupled with the phase-field model to investigate fracture behaviour. Material is assumed to be isotropic, therefore the total mechanical energy is considered for bulk degradation. Experiments are conducted on the samples aged for different humid conditions to understand tear strength. These test results are used to identify the phase-field damage material parameters from numerical investigation.
	\end{abstract}
\end{frontmatter}
\section{Introduction}
Crosslinked network polymer adhesives are becoming increasingly dominant in bonding technology in manufacturing of the lightweight material. Epoxy and polyurethane adhesives are the most popular chemically crosslinked network polymer adhesives. The present work focuses on polyurethane structural adhesives. These structural adhesives exhibit largely non-linear rate-dependent behaviour due to simultaneous elastic and viscous behaviour. Additionally, polyurethane adhesives are sensitive to the moisture of the environmental conditions due to hygroscopic properties. Hygroscopic behaviour leads to the diffusion of moisture from the environment causing material to age. The ageing of material is classified into chemical and physical ageing. Chemical ageing is an irreversible process due to the chemical interaction between moisture and polymer network resulting in breakage of bonds and forming new bonds. Whereas, physical ageing is reversible since there are no chemical interactions that influence the physical properties of the material over time. The primary focus of the present work is to formulate a material model to investigate the physical ageing of polyurethane adhesive under the influence of moisture.

Material models based on finite-strain viscoelastic mechanical behaviour are modelled primarily for rubber materials \cite{Haupt202}. These material models are classified into phenomenological and micro-mechanical models. Phenomenological continuum mechanical models are formulated based on invariants or principal strains \cite{Mooney,Rivlin1948,Ogden1972,Treloar4}. Micro-mechanical models are modelled based on the statistical polymer network theory \cite{Kuhn1936,Kuhn1942,Arruda,Miehe2004,Miehe2005}. This article utilizes phenomenological formulations to describe the finite-strain viscoelastic behaviour. Herein, a nearly incompressible behaviour is assumed with large deformations, therefore the deformation is decomposed into volumetric and isochoric parts \cite{Lubliner,HARTMANN20021439}. The decomposition provides a clear description of the physical behaviour of volume and shape-changing parts. Further, the isochoric part is decomposed into elastic and inelastic parts to describe rate-dependent behaviour using continuum-based rheological models. 

Modelling of crack propagation is an existing challenge in polymer materials \cite{Grellmann,Williams1987}. In this context, the crack propagation is well understood within the framework of theoretical continuum mechanics \cite{Krajcinovic}. The energy balance at the crack propagation boundary is described based on Griffith's criterion. Griffith's theory states that a crack propagates when the energy release rate at the crack propagation zone is higher than the surface energy built up. The conventional method in modelling crack separates the material into a broken and intact material by an interface. However, such a method requires a priori knowledge of the exact position of the interface and is complex to model in three-dimensional systems. Therefore, the phase-field method is developed to have a decisive advantage over sharp interface models since the explicit interface tracking becomes redundant \cite{ma14081913}.

A distinction is made between physical and mechanical approaches in modelling the phase-field material models. The physical model approaches are based on the Ginzburg-Landau phase transformation. In contrast, the mechanical approaches are based on Griffith's failure theory. A review of the different approaches in modelling phase-field ductile fracture is detailed by Ambati et al. \cite{Ambati2015}. These models use order parameters to distinguish broken and intact material by minimising the system's free energy \cite{Karma2001}. Phase-field fracture models describe crack propagation in homogeneous materials under different loads \cite{Bourdin2008,KUHN2010,Miehe2010}, including plastic effects \cite{Ambati2015,Hofacker2012,DUDA2015,AMBATI2016} and multi-physics problems \cite{MIEHE2015486,MIEHE2015,Spatschek}. Based on Griffith's theory, a model with position-dependent crack resistance was presented by Hossain et al. \cite{Hossain} for studies of fracture strength in materials.
\section{Finite-strain viscoelasticity}
\begin{figure}[H]
	\centering
	\vspace{4mm}
	\def\svgwidth{0.7\textwidth}
	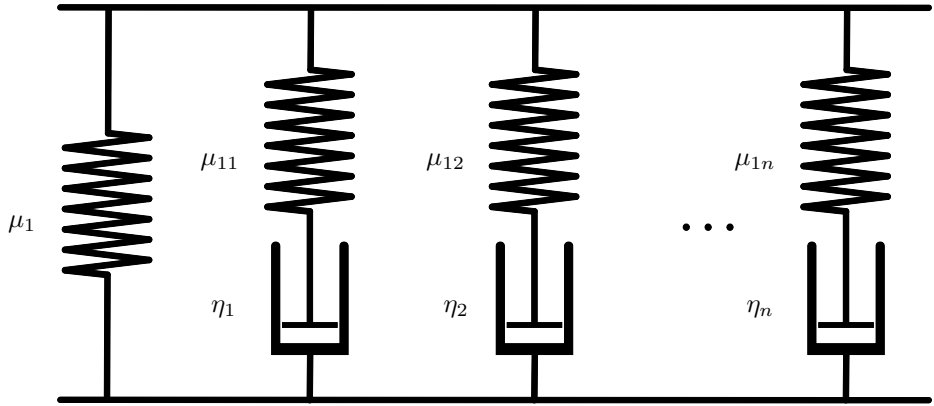
	\vspace*{2mm}	
	\caption{Rheological model of the viscoelasticity with $n$ Maxwell elements.}
	\label{finviskoelast40}
\end{figure}

A nearly incompressible material behaviour motivates the multiplicative decomposition of the deformation gradient tensor into its isochoric and volumetric components. This decomposition of the deformation gradient $\mathbf{F}$ is 
\begin{equation}
	\mathbf{F}=\mathbf{F}_{\rm vol}\cdot \bar{\TF},
	\label{defogr}
\end{equation}
where $\mathbf{F}_{\rm vol}$ and $\bar{\TF}$ are the volumetric and isochoric components respectively. The deformation tensor is decomposed into the elastic $\mathbf{F}_e$ and inelastic $\mathbf{F}_i$ parts \cite{Lee,LeeLiu,Govindjee1997,Lion,Lubliner,Reese,Tallec,Lubarda2004}. The decomposition introduces a fictitious intermediate configuration to represent rate-dependent behaviour. Each Maxwell element is introduced with a fictitious intermediate configuration. The decomposition of the deformation gradient is
\begin{equation}
	\TF\,=\,\TF_e\,\cdot\,\TF_i\,.
	\label{finvisko10}
\end{equation}
The elastic and inelastic components deformation gradient of equation \eqref{finvisko10} are enforced onto the isochoric component of the deformation gradient, thus leading to
\begin{equation}
	\bar{\TF}_e = \left(\det\TF_e\right)^{1/3}\TF_e, \hspace{3mm} {\rm and} \hspace{3mm}\bar{\TF}_i = \left(\det\TF_i\right)^{1/3}\TF_i.
\end{equation}
Consequently, the associated unimodular Cauchy-Green deformation tensors are
\begin{equation}
	\begin{aligned}
		& \bar{\TC}_e=J^{-2/3}\left(\bar{\TF}_e\right)^T \cdot \bar{\TF}_e;\,\hspace{11mm} \bar{\TC}_i^j=J^{-2/3}\left(\bar{\TF}_i^j\right)^T \cdot \bar{\TF}_i^j\\
		& \bar{\TB}_e=J^{-2/3}\left(\bar{\TF}_e\right)^{-T} \cdot \left(\bar{\TF}_e\right)^{-1};\,\hspace{3mm} \bar{\TB}_i^j=J^{-2/3}\left(\bar{\TF}_i^j\right)^{-T} \cdot \left(\bar{\TF}_i^j\right)^{-1}
	\end{aligned}
\end{equation}
The following relationship applies between the quantities
\begin{equation}
	\bar{\TB}_e= \bar{\TF}\cdot\left(\bar{\TC}_i^j\right)^{-1}\cdot \bar{\TF}
\end{equation}
The time derivative of inelastic Cauchy-Green deformation yields the rate of the inelastic right Cauchy-Green deformation tensor on the reference configuration as follows
\begin{equation}
	\dot{\bar{\TC}}_i^j = J^{-2/3}\left(\left(\bar{\TF}_i^j\right)^T \cdot \bar{\TF}_i^j\right)^\cdotp
\end{equation}
The inelastic right Cauchy-Green deformation tensor is regarded as an internal variable in the subsequent continuum mechanical description of the material behaviour. Its rate of inelastic right Cauchy-Green deformation tensor is described by evolution equation  \cite{Hauptpeter,Lionphd,Seldan}
\begin{equation}
	\begin{aligned}
		\dot{\bar{\TC}}_i^j\,&=\,\frac{4}{r_j} \left[\bar{\TC} - \frac{1}{3}{\rm{tr}}\left(\bar{\TC}\cdot\left(\bar{\TC}_i^j\right)^{-1} \right)\bar{\TC}_i^j\right]\\
	\end{aligned}
	\label{inelasticright}
\end{equation}
where $r_j$ is the relaxation time associated with individual Maxwell elements. The evolution equation is an outcome of the dissipation inequality described in the Appendix \ref{thermo}. The equation is solved by an implicit Euler method in time in combination with a local Newton method in space at each Gaussian point in the framework of the boundary value problem. ${\rm I}_1$ and ${\rm I}_3$ are the first and third invariants of the Cauchy-Green deformation tensor. These invariants are calculated as follows
\begin{equation}
	{\rm I}_1 = {\rm tr}\left(\mathbf{B}\right);\hspace{5mm}
	{\rm I}_3 = {\rm det}\left(\mathbf{C}\right) = {\rm det}\left(\mathbf{B}\right) = J^2,
\end{equation}
and the modified counterparts of the invariants are calculated as 
\begin{equation}
	\bar{{\rm I}}_1 = J^{-2/3} {\rm I}_1;\hspace{5mm} \bar{{\rm I}}_3={\rm det}\left(\bar{\mathbf{C}}\right) = {\rm det}\left(\bar{\mathbf{B}}\right).
\end{equation}
The viscoelastic material model is based on a continuum mechanical description motivated by a rheological model. Rheological models are described by combining the spring element with the Maxwell element in parallel to model relaxation behaviour. The discrete spectrum of relaxation time is considered by combining the spring element with several Maxwell elements as shown in Figure \ref{finviskoelast40}. 
A nearly incompressible deformation is assumed in modelling the non-linear viscoelastic behaviour. Therefore, an uncoupled response of free energy \cite{HARTMANN20021439,SIMO1987153} is used in the formulation. Total mechanical energy of the rheological model consisting of $j = 1,2, . . . , n$ is
\begin{equation}
	W_0\left(J, {\rm I}_1^{\bar{\mathbf{B}}},{\rm I}_1^{{\bar{\mathbf{B}}}^j_e} \right) = W_{\rm vol} \left(J \right)+ W_{\rm eq} \left(\bar{\rm I}_1^{\bar{\mathbf{B}}} \right) +  \sum_{j=1}^{n} W^j_{\rm neq}\left({\rm I}_1^{\bar{\mathbf{B}}^j_e}\right).
	\label{finvisko450-1}
\end{equation}
The equilibrium part of the free energy function is motivated by a polynomial function of the isochoric first invariant of the left Cauchy-Green deformation tensor $\bar{\rm I}_1^{\bar{\mathbf{B}}}$ using the Yeoh model \cite{Yeoh,Yeoh1993} 
\begin{equation}
	W_{\rm eq}\left(\bar{\rm I}_1^{\bar{\mathbf{B}}}\right) = c_{10}\!\left(\bar{\rm I}_1^{\bar{\mathbf{B}}}-3\right) +\, c_{20}\!\left(\bar{\rm I}_1^{\bar{\mathbf{B}}}-3\right)^2\! +\, c_{30}\!\left(\bar{\rm I}_1^{\bar{\mathbf{B}}}-3\right)^3\!\!\!\!,
	\label{isofree}
\end{equation}
where $c_{10},\,c_{20}\, {\rm and}\, c_{30}$ are the stiffness parameters. A general quadratic form is considered in current formulation \cite{Holzapfel,Brink1996,ODEN1978445,Ogden1972,Ogden1985} for the volumetric part of free energy density function 
\begin{equation}
	W_{\rm vol}\left(J\right) = \frac{1}{G}\left(J-1\right)^2, 
	\label{volfree}
\end{equation}
where $G$ is the compression modulus. The non-equilibrium free energy for $j=1,2,...,n$ Maxwell elements is computed as the sum of the individual energies of Maxwell elements
\begin{equation}
	\sum_{j=1}^{n} W^j_{\rm neq}\left({\rm I}_1^{{\mathbf{B}}^j_e}\right)=\sum_{j=1}^{n} W_{\rm neq}^j\left(\bar{\rm I}_1^{\bar{\mathbf{B}}^j_e}\right) = \sum_{j=1}^{n} c_{10j}\left(\bar{\rm I}_1^{\bar{\mathbf{B}}^j_e}-3\right).
	\label{nonelasenerg}
\end{equation}
The corresponding constitutive equation for stresses is calculated as the sum of volumetric, equilibrium and $j=1,2,...,n$ non-equilibrium stress components
\begin{equation}
	\TT\left(J, {\rm I}_1^{\bar{\mathbf{B}}},{\rm I}_1^{{\bar{\mathbf{B}}}^j_e} \right) = \TT_{\rm vol} \left(J \right)+ \TT_{\rm eq} \left(\bar{\rm I}_1^{\bar{\mathbf{B}}} \right) +  \sum_{j=1}^{n} \TT^j_{\rm neq}\left({\rm I}_1^{\bar{\mathbf{B}}^j_e}\right).
\end{equation}
\section{Phase-field damage}
This section describes the phase-field damage model used to investigate fracture in the materials that exhibit rate-dependent behaviour due to viscoelastic properties with large deformations. The basic idea behind the variational formulation of the phase-field fracture model is to minimise the free energy by obeying a kinematically admissible displacement field. The free energy $W$ is based on Francfort-Marigo functional \cite{Francfort1998} to describe crack follows  
\begin{equation}
	W \left(\bar{\rm I}_1^{\bar{\mathbf{B}}}, \bar{\rm I}_1^{\bar{\mathbf{B}}^j_e},J,\phi \right) = W_b \left(\bar{\rm I}_1^{\bar{\mathbf{B}}}, \bar{\rm I}_1^{\bar{\mathbf{B}}^j_e},J,\phi \right) + W_s\left(\phi\right),
	\label{eq:toten}
\end{equation}
where $W_b$ and $W_s$ are the bulk and surface energies. The surface energy is given by Griffith's theory \cite{Griffith} to predict crack initiation and branching
\begin{equation}
	W_s\left(\phi\right) = \int\limits_\Gamma E_c {\rm d}\Gamma.
\end{equation}
$E_c$ is the critical energy release rate to describe the crack resistance of the material. The surface energy is regularised with the crack density functional $\gamma$ \cite{Bourdin2000} to obtain volume integral
\begin{equation}
	W_s(\phi) = \int\limits_{\Omega} E_c \gamma(\phi, \grad\, \phi)\,{\rm dV}.
	\label{eq:phasefun}
\end{equation}
The crack energy density functional introduces phase-field variable $\phi(\Tx)\in \left[0,1\right]$ to distinguish between intact $\phi(\Tx)=1$ and cracked $\phi(\Tx)=0$ material. The crack surface density is introduced with a second-order regularised function as 
\begin{equation}
	\begin{aligned}
		\gamma\left(\phi, \, \grad\,\phi\right)=  \left[\frac{1}{2\ell_f} \left(1-\phi\right)^2 + \frac{\ell_f}{2}|\grad\,\phi|^2\right],
		\label{elliptical}
	\end{aligned}
\end{equation}
wherein, $\ell_f$ is the length scale parameter introduced to control the size of the crack zone. The mechanical energy stored in the bulk degrades as the crack propagates with time. A degradation function is introduced to consider the degradation of bulk energy $W_b$ as: 
\begin{equation}
	W_b \left(\bar{\rm I}_1^{\bar{\mathbf{B}}}, \bar{\rm I}_1^{\bar{\mathbf{B}}^j_e},J,\phi \right)  = g(\phi)W_0 \left(\bar{\rm I}_1^{\bar{\mathbf{B}}}, \bar{\rm I}_1^{\bar{\mathbf{B}}^j_e},J \right),
	\label{eq:bulk}
\end{equation}
where $g(\phi)$ is the degradation function and $W_0$ is the viscoelastic free energy defined in equation \eqref{finvisko450-1}. The degradation function plays a vital role in interpolating stresses to characterise the intact and broken state of the material. The degradation of bulk energy for intact and broken state defined by phase-field variable $\phi\in\left[0,\, 1\right]$ have to satisfy conditions
\begin{equation}
	\begin{aligned}
		&W_b \left(\bar{\rm I}_1^{\bar{\mathbf{B}}}, \bar{\rm I}_1^{\bar{\mathbf{B}}^j_e},J,\phi=1 \right) = W_0 \left(\bar{\rm I}_1^{\bar{\mathbf{B}}}, \bar{\rm I}_1^{\bar{\mathbf{B}}^j_e},J \right),\,\hspace{1mm}
		W_b \left(\bar{\rm I}_1^{\bar{\mathbf{B}}}, \bar{\rm I}_1^{\bar{\mathbf{B}}^j_e},J,\phi=0 \right) = 0\, ,\\
		&\partial W_b \left(\bar{\rm I}_1^{\bar{\mathbf{B}}}, \bar{\rm I}_1^{\bar{\mathbf{B}}^j_e},J,\phi \right) < 1\,\,\hspace{2mm} {\rm and}\,\,\hspace{2mm}
		\partial W_b \left(\bar{\rm I}_1^{\bar{\mathbf{B}}}, \bar{\rm I}_1^{\bar{\mathbf{B}}^j_e},J,\phi=0 \right) = 0.
	\end{aligned}
\end{equation}
Equations \eqref{eq:phasefun}, \eqref{elliptical} and \eqref{eq:bulk} are substituted in the equation \eqref{eq:toten} and integrated over the volume to derive the free energy density  \cite{Bourdin2008,Francfort1998} 
\begin{equation}
	\begin{aligned}
		W \left(\bar{\rm I}_1^{\bar{\mathbf{B}}}, \bar{\rm I}_1^{\bar{\mathbf{B}}^j_e},J,\phi \right)= & \int\limits_{\Omega} g(\phi) W_0 \left(\bar{\rm I}_1^{\bar{\mathbf{B}}}, \bar{\rm I}_1^{\bar{\mathbf{B}}^j_e},J \right) {\rm dV}\,\, +
		\int\limits_{\Omega} E_c \left[\frac{1}{2\ell_f} \left(1-\phi\right)^2 + \frac{\ell_f}{2}|\grad\,\phi|^2\right]{\rm dV}.
	\end{aligned}
	\label{eq:totene}
\end{equation} 
In this article, a monotonically decreasing function is considered to describe the decay of the bulk energy. A second-order degradation function is considered with an additional regularisation parameter $\zeta$ \cite{Bourdin2000,Heister2015,Kuhn2015} to interpolate the bulk energy  
\begin{equation}
	g(\phi) = (1-\zeta)\phi^{2}+\zeta.
	\label{eq:degfun}
\end{equation}
The regularisation parameter $\zeta>0$ is employed to guarantee a converged solution. A smaller value is selected to avoid overestimation of mechanical energy \cite{BORDEN201277,Bourdin2000,Braides2000}. The energy degradation function satisfies the condition
\begin{equation}
	g(\phi=0) = 0, \, g(\phi=1) = 1\, {\rm and}\, g'(\phi=0) = 0,
\end{equation}
where $g(\phi=0) = 0$ damaged material, $g(\phi=1) = 1$ describes the intact material and $g'(\phi=0) = 0$ controls the stored mechanical energy in the phase-field evolution equation.
\section{Governing balance equations}
The weak form of the free energy function is derived by applying the variational principle to the total potential energy with the field variables $\left(\mathbf{u},\phi\right)$ 
\begin{equation}
	\delta W = \left(\frac{\partial W}{\partial \mathbf{u}}\right)\colon \delta \mathbf{u} + \left(\frac{\partial W}{\partial \phi}\right)\colon\delta\phi.
\end{equation}
Furthermore, the continuum domain $\Omega$ is integrated over the total volume ${\rm dV}$ leading to the weak form for the admissible test functions of phase-field $\delta\phi$ and displacement field $\delta\mathbf{u}$ 
\begin{equation}
	\begin{aligned}
		\delta W = \int\limits_{\Omega} &\left\{g(\phi)\, \TT \colon{\rm grad ^s}\delta\mathbf{u} + g'(\phi)\, \delta\phi W_0\right\}{\rm dV} + 
		& \int\limits_{\Omega} \left\{E_c \left[-\frac{1}{\ell_f} \left(1-\phi\right) \delta\phi + \ell_f \,\grad\, \phi\,  \grad \delta\phi\right]\right\}{\rm dV},
		\label{eq:variational}
	\end{aligned}
\end{equation}
${\rm grad ^s}\delta\mathbf{u} = \frac{1}{2}\left[{\rm grad}\delta\mathbf{u} + \left({\rm grad}\delta\mathbf{u}\right)^T\right]$ is involved due to symmetric stress tensor. After substituting the degradation function defined in the equation \eqref{eq:degfun} and the derivative of degradation function $g'(\phi) ={\partial g(\phi)}/{\partial\phi}$ in the coupled form given in equation \eqref{eq:variational} follows: 
\begin{equation}
		\begin{aligned}
			\delta W = \int\limits_{\Omega}& \left\{\left[(1-\zeta)\phi^{2}+\zeta\right] \TT\colon{\rm grad ^s}\delta\mathbf{u}\right\}{\rm dV} + \\
			\int\limits_{\Omega}& \left\{2\left(1-\zeta\right) \phi\, \delta\phi W_0 + E_c \left[-\frac{1}{\ell_f} \left(1-\phi\right) \delta\phi + \ell_f\, \grad\, \phi\,  \grad \delta\phi\right]\right\}{\rm dV}.
		\end{aligned}
	\label{pfds}
\end{equation}
The strong form of the coupled formulation gives the local statement for the phase-field method and is derived by applying the divergence principle to equation \eqref{pfds}
\begin{subequations}
	\begin{align}
		{\rm div}\left(\left[(1-\zeta)\phi^{2}+\zeta\right] \TT\right) &= \mathbf{0} \label{eq:mech}\\
		\underbrace{2\left(1-\zeta\right) \phi\, W_0}_{\rm driving\, force} + \underbrace{E_c \left[-\frac{1}{\ell_f} \left(1-\phi\right) + \ell_f \,\div \phi \right]}_{\rm resistance\, to\, crack} &= 0.
		\label{eq:phase}
	\end{align}
\end{subequations}
Equation \eqref{eq:mech} refers to the balance of momentum describing the viscoelastic response, and equation \eqref{eq:phase} is the phase-field evolution of the diffusive crack. The first term of the phase-field evolution is responsible for driving the crack, and the second term refers to the geometric resistance to the propagation of the crack. $W_0$ is the energy stored in material domain with $W_0 =\max \limits_{0<\phi<t} W_0\left( \mathbf{x}, \phi\right) $ to avoid an irreversibility in the crack propagation \cite{MIEHE2015486}.
\subsection{Finite element implementation}
It is convenient to express the partial differential equations \eqref{eq:mech} and \eqref{eq:phase} in their weak forms to develop a numerical solution scheme using finite element method
\begin{equation} \label{eq:15}
		\begin{aligned}
			\mathbf{r}^{\mathbf{u}}_{i} = \int\limits_{\Omega} \left\{\left[(1-\zeta)\phi^{2}+\zeta\right] \boldsymbol{ \TT} \colon{\rm grad ^s}\delta\mathbf{u}\right\}{\rm dV} &= \mathbf{0}\\
			{\rm r}^{\phi}_{i} = \!\!\int\limits_{\Omega}\!\! \left\{2\left(1-\zeta\right) \phi\, \delta\phi W_0 \!+\! E_c \left[-\frac{1}{\ell_f} \left(1-\phi\right) \delta\phi \!+\! \ell_f\, \grad \phi \, \grad \delta\phi\right]\right\}{\rm dV} &= 0.
		\end{aligned}
\end{equation}
In this context, the displacement $\mathbf{u}$ and phase-field variable $\phi$ are discretised in space as 
\begin{equation}\label{expres}
	\mathbf{u} = \sum_{i=1}^{{\rm n}_{\rm ele}} \mathbf{N}_i^{\mathbf{u}} \mathbf{u}_i \hspace{1 cm} \phi = \sum_{i=1}^{{\rm n}_{\rm ele}} {N}_i^\phi {\phi}_i,
\end{equation}
where ${N}_i^\phi$ is the shape function concerning the phase-field variable and $\mathbf{N}_i^{\mathbf{u}}$ is the displacement shape function used to interpolate between the quantities at the quadrature points. The displacement shape function in three dimensions is given by: 
\begin{equation} \label{eq:18}
	\mathbf{N}_i^{\mathbf{u}} = \begin{bmatrix} 
		{\rm N}_i & 0 & 0 \\ 
		0 & {\rm N}_i & 0 \\ 
		0 & 0 & {\rm N}_i\\ 
	\end{bmatrix}.
\end{equation}
In equation \eqref{eq:18} $N_i$ is the value of the shape function of the displacement field $\mathbf{u}_i = \left({\rm u}_x,{\rm u}_y, {\rm u}_z\right)^T$ at the quadrature points associated with the respective nodes. The gradient of the phase-field variable $\phi_i$ follow 
\begin{equation} \label{eq:19}
	\grad \phi = \sum_{i=1}^{{\rm n}_{\rm ele}} {\TS}_i^{\phi} {\phi}_i.
\end{equation}
Herein, the $\mathbf{S}$ matrix is introduced as
\begin{equation} \label{eq:20}
	\mathbf{S}_i^{\mathbf{u}} = \begin{bmatrix} 
		{\rm N}_{i,x} & 0 & 0\\ 
		0 & {\rm N}_{i,y} & 0\\ 
		0 & 0 & {\rm N}_{i,z}\\
		{\rm N}_{i,y} & {\rm N}_{i,x} & 0\\
		0 & {\rm N}_{i,z} & {\rm N}_{i,y} \\
		{\rm N}_{i,z} & 0 & {\rm N}_{i,x}\\
	\end{bmatrix} \hspace{1 cm}
	\mathbf{S}_i^{\phi} = \begin{bmatrix} 
		{\rm N}_{i,x}\\ 
		{\rm N}_{i,y} \\ 
		{\rm N}_{i,z} \\
	\end{bmatrix},
\end{equation}
where, $N_{i,x}$, $N_{i,y}$ and $N_{i,z}$ are the derivatives of the shape functions evaluated as $\partial N_i/\partial x$, $\partial N_i/\partial y$ and $\partial N_i/\partial z$. In the same way, virtual quantities of the displacement and phase-field variables are approximated as 
\begin{equation} \label{eq:21}
	\begin{split}
		\delta \mathbf{u} = \sum_{i=1}^{\text{n}_\text{ele}} \mathbf{N}_i^{\mathbf{u}} \delta \mathbf{u}_i \hspace{1.7 cm} \delta \phi = \sum_{i=1}^{\text{n}_\text{ele}} {N}_i^\phi \delta {\phi}_i \\
		{\rm grad^s}\delta\mathbf{u} = \sum_{i=1}^{\text{n}_\text{ele}} 
		\mathbf{S}_i^{\delta \mathbf{u}} \mathbf{u}_i \hspace{1 cm} \grad\, \delta \phi = \sum_{i=1}^{\text{n}_\text{ele}} {\TS}_i^{\phi} {\delta \phi}_i.
	\end{split}
\end{equation}
The coupled system of equations is non-linear, therefore the coupled problem is solved iteratively using the Newton-Raphson method. The finite element formulation to solve the coupled system of equations with an incremental method follows: 
\begin{equation}\label{staggequa}
	{\begin{bmatrix} 
			\mathbf{{K}}^\mathbf{uu} & \mathbf{{K}}^{\mathbf{u}\phi} \\ 
			\mathbf{{K}}^{\phi\mathbf{u}} & \mathbf{{K}}^{\phi \phi} \\ 
	\end{bmatrix}} 
	{\begin{bmatrix} d \mathbf{\mathbf{u}}\\ 
			d {{{\phi}}} \\ 
	\end{bmatrix}} = {\begin{bmatrix} -\mathbf{{r}}_{\mathbf{u}} (\mathbf{u}_{\textrm{i}}) \\ 
			-{{\rm r}}_{\phi} ({\phi}_{\textrm{i}})
	\end{bmatrix}}.
\end{equation}
Since the primary variables defined in the equation \eqref{expres} hold for the arbitrary values $\delta \Tu$ and $\delta \phi$, the residuals of the coupled system of equations defined in equation \eqref{eq:15} are expressed in term of the virtual quantities given with equation \eqref{eq:21} as 
\begin{equation}
		\begin{aligned}
			\mathbf{r}^{\mathbf{u}}_{i} = \int\limits_{\Omega} \left\{\left[(1-\zeta)\phi^{2}+\zeta\right] \boldsymbol{\TT} \colon(\mathbf{S}_i^{\mathbf{u}})^\text{T}\right\}{\rm dV} &= \mathbf{0},\\
			{\rm r}^{\phi}_{i} = \int\limits_{\Omega}\left\{2\left(1-\zeta\right) \phi\, N_i W_0 + E_c \left[-\frac{1}{\ell_f} \left(1-\phi\right) N_i + \ell_f ( \mathbf{S}_i^{\phi})^{T} \mathbf{S}_j^{\phi}\right]\right\}{\rm dV} &= {0},
		\end{aligned}
\end{equation}
and the elements of the tangent matrix are
\begin{equation}
		\begin{aligned}
			&\mathbf{K}_{i,j}^{\mathbf{uu}} = \frac{\partial \mathbf{r}_i^\mathbf{u}}{\partial \mathbf{u}_j} = \int\limits_{\Omega}  \left( \left( 1 - \zeta \right) \phi^2 + \zeta \right)\left(\mathbf{S}^{\mathbf{u}}_i \colon\TenF{\kapp}\colon \mathbf{S}^{\mathbf{u}}_j +  \mathbf{S}^{\mathbf{u}}_i \colon \left[\boldsymbol{\TT}\cdot\mathbf{S}^{\mathbf{u}}_j\right]\right) {\rm dV}, \\
			&\mathbf{K}_{i,j}^{\mathbf{u}\phi} = \frac{\partial \mathbf{r}_i^\mathbf{u}}{\partial \phi_j} = \int\limits_{\Omega} 2(1-\zeta)\,\phi\, \mathbf{S}^{\mathbf{u}}_i\colon \boldsymbol{\TT}^T \mathbf{N}_j^{\mathbf{u}}\, {\rm dV},\\
			&\mathbf{K}_{i,j}^{\phi\mathbf{u}} = \frac{\partial \mathbf{r}_i^{\phi}}{\partial \mathbf{u}_j} = \int\limits_{\Omega} 2(1-\zeta)\,\phi\, \mathbf{N}_i^{\mathbf{u}}\, \boldsymbol{\TT}^T \colon\mathbf{S}^{\mathbf{u}}_j\, {\rm dV},\\
			&\mathbf{K}_{i,j}^{\phi \phi} = \frac{\partial \mathbf{r}_i^{\phi}}{\partial \phi_j} = \int\limits_{\Omega} \left\lbrace  \left(1 - \zeta \right) W_0\, {\rm N}_i^\phi\, {\rm N}_j^\phi + E_c \left[\frac{1}{\ell_f} {\rm N}_i^\phi\, {\rm N}_j^\phi + \ell_f (\mathbf{S}^{\phi}_i)^{\text{T}}\colon \mathbf{S}^{\phi}_j \right]\right\rbrace {\rm dV}. \\
		\end{aligned}
\end{equation}
The coupled system of equations is solved simultaneously using a monolithic approach with Newton's iterative method. 
\subsection{Boundary conditions}
The boundary conditions are postulated for the displacement field variable $\mathbf{u}$ and the phase-field damage variable $\phi$ to solve the phase-field damage formulation. To this end, the surface $\partial\Omega$ is decomposed to the primary fields, the displacement and damage fields 
\begin{equation}
	\partial\Omega=\partial\Omega_{\mathbf{u}}^D\cup\partial\Omega_{\Tt}^N\,\, {\rm and}\,\, \partial\Omega=\partial\Omega_{\phi}^D\cup\partial\Omega_{\nabla\phi}^N
\end{equation}
with $\partial\Omega_{\mathbf{u}}^D\cap\partial\Omega_{\mathbf{u}}^N = \varnothing$ and $\partial\Omega_{\phi}^D\cap\partial\Omega_{\nabla\phi}^N = \varnothing$. The prescribed displacement $\mathbf{u}$ and traction $\Tt$ of the mechanical problem on the boundaries are postulated with the Dirichlet and Neumann boundary conditions  
\begin{equation}
	\mathbf{u}\left(\mathbf{x},t\right) = \mathbf{u}_D\left(\mathbf{x},t\right) \,\, {\rm on} \,\,\partial\Omega_{\Tu}^D \,\, {\rm and} \,\, \TT\cdot\Tn = \Tt \,\, {\rm on} \,\,\partial\Omega_{\Tt}^N.
\end{equation}
For the phase-field damage, the cracked region is constrained by the Dirichlet and the Neumann boundary conditions on the crack surface with 
\begin{equation}
	\phi\left(\mathbf{x},t\right) = 0 \hspace{2mm} \text{at} \hspace{2mm} \mathbf{x} \in \partial\Omega_{\phi}^D\,\,{\rm and}\,\,\nabla\phi\cdot\Tn=0\,\,{\rm on}\,\,\partial\Omega_{\nabla\phi}^N.
\end{equation}
\section{Numerical investigation}
\begin{figure}[H]
	\centering
	\def\svgwidth{0.7\textwidth}
	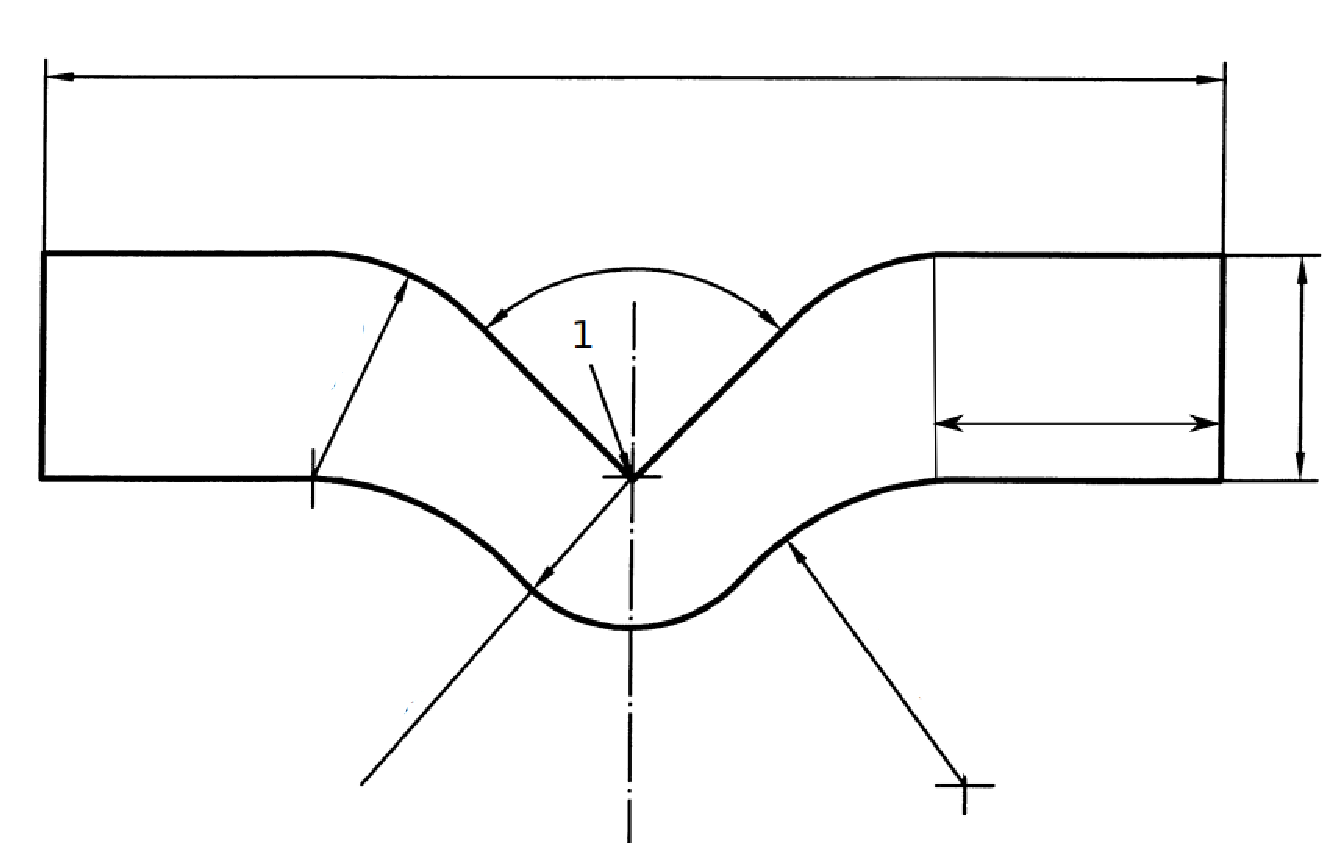
	\caption{Geometry of the angle sample based on DIN ISO 34-1: all dimensions are in millimetres}
	\label{angular}
\end{figure}

\begin{figure}[H]
	\centering
	\def\svgwidth{0.7\textwidth}
	\input{phasefield/energy.tex}
	\caption{Critical fracture energy release rate $E_c$ of adhesive-A samples aged at different relative humidities (r.H.) in the atmosphere at $60^{\circ}\,\rm C$}
	\label{energy}
\end{figure}
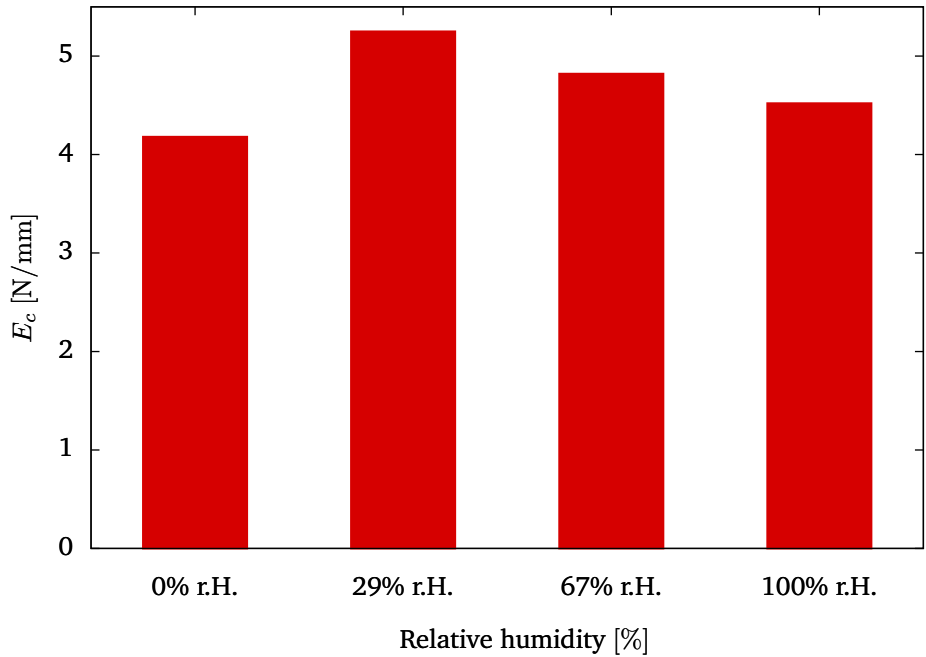

The experimental investigation performed on the commercial adhesive-A in project IGF-project 19730 N \cite{Josyula} is used in the present work for numerical investigation. The tensile tests performed on the aged samples are used to identify the viscoelastic parameters. The aged samples are prepared at four different relative humid conditions $0\%$r.H, $29\%$r.H, $67\%$r.H, and $100\%$r.H at an isothermal condition of 60° C. The viscoelastic model is formulated by using a spring element connected in parallel to the four Maxwell elements. The relaxation times of the Maxwell elements are assumed constant irrespective of ageing conditions. The assumption of constant relaxation time is analogous to the viscoelastic behaviour proposed in \cite{Goldschmidt2018,Johlitz2012} to fit the loading rates used in the experimental investigation. The stiffness parameters of the polyurethane adhesive were identified with the curve fitting process using Matlab optimization toolbox. The identified parameters of the viscoelastic material for all the investigated aged samples are listed in Table \ref{table1}.
\begin{table}[H]
	\caption{Material parameters of the finite-strain viscoelastic material model identified for the different relative humid atmospheres at $60^{\circ}\,\rm C$}
	\small\addtolength{\tabcolsep}{-3.5pt}
	\renewcommand{\arraystretch}{1.7}
	\centering
	\begin{tabular}{|c|c|c|c|c|c|c|}
		\hline
		\multicolumn{7}{|c|}{Material parameters of finite-strain viscoelastic model}                                                                                                    \\ \hline
		&                  & \begin{tabular}[c]{@{}c@{}}Relaxation\\  times {[}s{]}\end{tabular} & $0\%$ r.H. & $29\%$ r.H. & $67\%$ r.H. & $100\%$ r.H. \\ \hline
		\multirow{4}{*}{Equilibrium}     & $c_{10}\,[\rm MPa]$  &                                                                     & 9.886      & 7.886       & 7.196       & 7.072        \\ \cline{2-7} 
		& $c_{20}\,[\rm MPa]$  &                                                                     & -1.414     & -1.357      & -1.122      & -1.128       \\ \cline{2-7} 
		& $c_{30}\,[\rm MPa]$  &                                                                     & 3.214      & 1.443       & 0.918       & 0.872        \\ \cline{2-7} 
		& $D\,[\rm MPa]$       &                                                                     & 0.306      & 0.244       & 0.241       & 0.314        \\ \hline
		\multirow{4}{*}{Non-equilibrium} & $c_{101}\,[\rm MPa]$ & 0.5                                                                 & 4.886      & 2.886       & 2.296       & 2.172        \\ \cline{2-7} 
		& $c_{102}\,[\rm MPa]$ & 10                                                                  & 0.886      & 0.231       & 0.139       & 0.107        \\ \cline{2-7} 
		& $c_{103}\,[\rm MPa]$ & 100                                                                 & 0.055      & 0.017       & 0.014       & 0.011        \\ \cline{2-7} 
		& $c_{104}\,[\rm MPa]$ & 1000                                                                & 0.005      & 0.003       & 0.002       & 0.001        \\ \hline
	\end{tabular}
	\label{table1}
\end{table} 
The primary focus of this work is to understand the damage behaviour in the polyurethane adhesive due to tensile load for samples aged under the influence of moisture. To this end, the sample preparation and experimental investigation is carried out by following DIN ISO 34-1. Angular specimens proposed in DIN ISO 34-1 (shown in Figure \ref{angular}) are prepared with a thickness of $2\,\rm mm$ to perform the tear test. Aged samples are manufactured in different humid conditions at an isothermal condition of $60^\circ\,\rm C$ to investigate the influence of moisture on the tear strength. Similar to tensile test samples, the angular specimens are aged under humid atmospheric conditions under $0\%$r.H, $29\%$r.H, $67\%$r.H, and $100\%$r.H condition at $60^\circ\,\rm C$. To perform a tear test an initial crack of $9.13\,\rm mm$ is imposed on the samples at the notch. Material is clamped approximately at $22\,\rm mm$ on both free ends are assumed to be rigid. Therefore, the clamped volume of material is not considered in the numerical investigation. 
For the numerical investigation, the CAD model is spatially discretized into the three-dimensional hexahedral mesh. The crack initiates and propagates at the cross-section of the notch until failure. Therefore, the mesh is refined at the transition zone satisfying the mesh refinement condition proposed by Miehe et al. \cite{Miehe2010}. The crack length-scale parameter is set to the length of the crack imparted in the sample as $\ell_f=9.31\,\rm mm$ and the length of mesh size of $h=1.23\,\rm mm$ is maintained to satisfy the mesh refinement condition. Finally, displacement boundary condition $u_y = 0.011\,\rm mm$ is applied to investigate the failure. The finite element model is solved monolithically using Newton's iterative method in a quasi-static process. The solution scheme is solved in several time steps with a constant time increment until fracture. The spatially discretized model is defined with the viscoelastic parameters listed in Table \ref{table1}. The critical energy release rate is an essential parameter required for the phase-field material model. This parameter is identified from the fitting force-displacement curve from numerical analysis with tear test using Matlab optimization toolbox \cite{Nelder}. The optimal critical energy release rate $E_c$ are listed in Table \ref{MPPFD} that are identified individually for the dry ($0\%$r.H) and aged samples (humid conditions: $29\%\,\rm r.H.$, $67\%\,\rm r.H.$ and $100\%\,\rm r.H.$) manufactured under the isothermal condition of 60 °C. 
\begin{table}[H]
	\centering
	\caption{Critical energy release rate identified for dry and aged samples prepared and tested at an isothermal condition of 60 °C} \label{MPPFD}
	\small\addtolength{\tabcolsep}{3pt}
	\renewcommand{\arraystretch}{2}
	\begin{tabular}{|c|c|c|c|c|}
		\hline
		\multicolumn{5}{|c|}{Identified critical energy release rate} \\ \hline
		ageing condition & $0\%$ r.H.  & $29\%$ r.H.       & $67\%$ r.H.       & $100\%$ r.H.      \\ \hline
		$E_c\,\left[\rm N/mm\right]$ & $4.18\,\rm N/mm$ & $5.25\,\rm N/mm$  & $4.82\,\rm N/mm$  & $4.52\,\rm N/mm$  \\ \hline
	\end{tabular}
\end{table}
The force-displacement data from the simulation are compared with the experimental test data to validate the identified optimal material parameters. The test used in the comparison is the mean value calculated from the series of three angular test samples manufactured at the same atmospheric conditions \cite{Josyula}. 
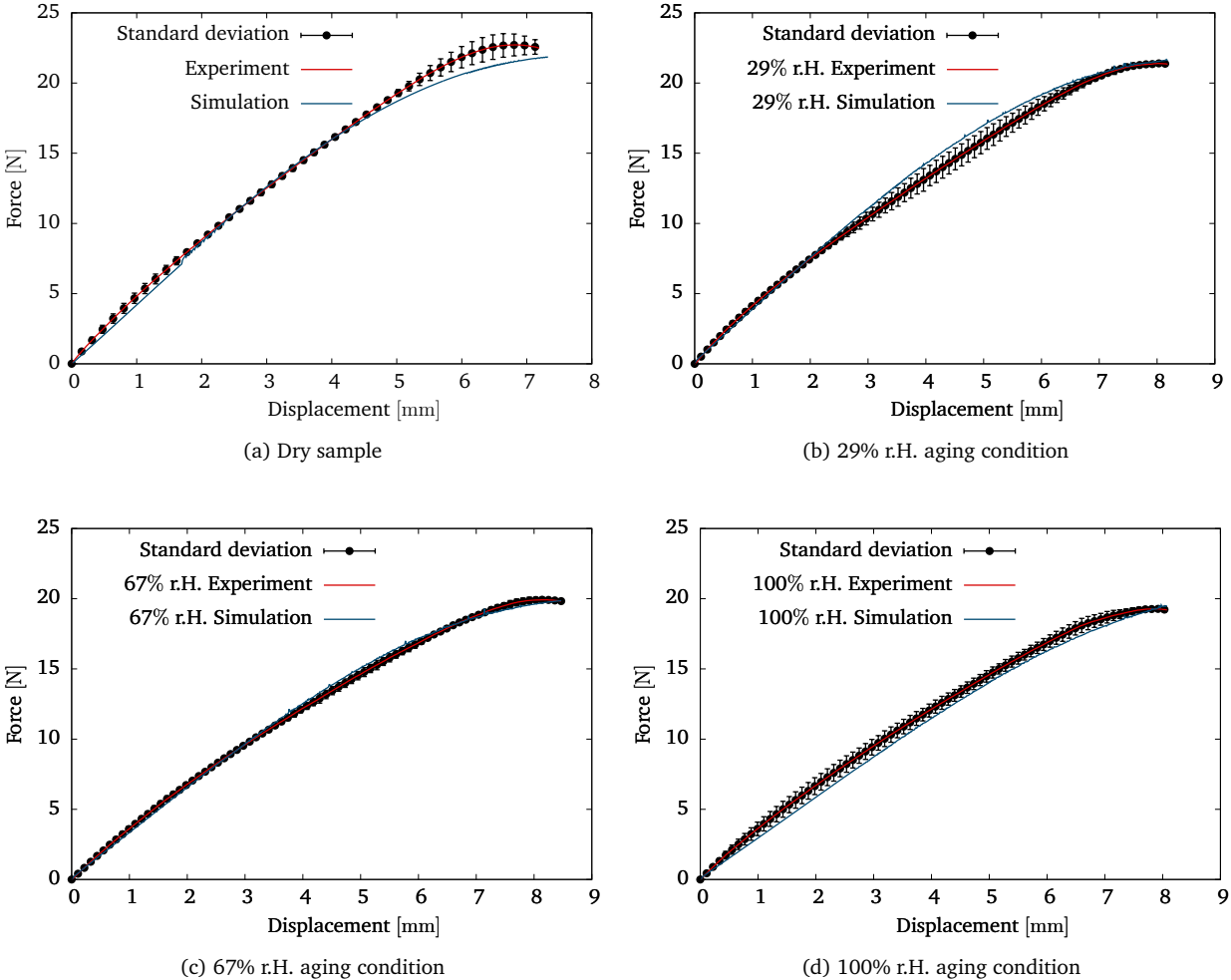
\begin{figure}[H]
	\begin{subfigure}{0.5\textwidth}
		\scalebox{0.65}{\input{phasefield/crack2VST}}
		\caption{Dry sample}
	\end{subfigure}%
	\hfill
	\begin{subfigure}{0.5\textwidth}
		\scalebox{0.65} {\input{phasefield/29RH_KVSD}}
		\caption{29\% r.H. aging condition}\label{29rhdama}
	\end{subfigure}\vspace{5mm}
	\begin{subfigure}{0.5\textwidth}
		\scalebox{0.65} {\input{phasefield/67RH_KVSD}}
		\caption{67\% r.H. aging condition}\label{67rhdama}
	\end{subfigure}
	\hfill
	\begin{subfigure}{0.5\textwidth}
		\scalebox{0.65} {\input{phasefield/100RH_KVSD}}
		\caption{100\% r.H. aging condition}\label{100rhdama}
	\end{subfigure}
	\caption{Tear tests are performed on angular samples aged under different humid conditions at $60^\circ$C}
	\label{orhdama}
\end{figure}
Figure \ref{orhdama} compares the experimental and the simulation data, where the deviation between the curves is shown with the error bars. The deviation in the form of small error bars indicates the problem is well posed, thus validating the identified material parameters. The critical energy release rate of the aged samples shown in Figure \ref{energy} indicates that the adhesive material becomes ductile under the influence of moisture, thus leading to an increase in the critical energy release rate. The critical energy release rate is minimum for the dry sample and reaches a maximum for the sample saturated at $29\%$ relative humidity.
\section{Summary}
In this article, the viscoelastic behaviour is modelled using on the rheological model based on continuum mechanics by combining elastic spring and Maxwell elements. Four Maxwell elements are used in combination with elastic spring elements in parallel and the energies are defined using phenomenological models. The viscoelastic model is coupled with the phase-field model to investigate the damage behaviour of the aged polyurethane adhesives under the influence of moisture. The phase-field damage model is a promising damage material model based on the Griffith fracture energy. These fracture models describe the crack evolution with a continuous order variable differentiating the intact and damaged material with diffusive crack.

The experimental investigations performed on commercial polyurethane adhesive called adhesive-A in IGF-project 19730 N \cite{Josyula} are used for numerical investigation. The fracture behaviour of the material is investigated under the influence of moisture at an isothermal condition. Aged samples are prepared for 29\% r.H., 67\% r.H. and 100\% r.H. condition at 60°C. Viscoelastic material model parameters are identified from the tensile test performed on the dry and aged samples. These viscoelastic parameters are used in the numerical investigation of the damage behaviour in the aged samples using the phase-field damage model. The tear strength on the samples is performed based on the DIN ISO 34-1 standard using angular specimens. The critical energy release rate is identified using an optimization algorithm with Matlab toolbox. The deviation between the experimental and simulation results is negligible validating the material parameters.


\section*{Acknowledgments}
The research project 19730 N "Berechnung des instation\"aren mechanischen Verhaltens von alternden Klebverbindungen unter Einfluss von Wasser auf den Klebstoff" of the research association Forschungsvereinigung Stahlanwendung e.V. (FOSTA), D\"usseldorf was supported by the Federal Ministry of Economic Affairs and Energy through the AiF as part of the program for promoting industrial cooperative research (IGF) on the basis of a decision by the German Bundestag. The experimental investigations are carried out by the project associates from Lehrstuthl f\"ur Angewandte Mechanik, Universit\"at des Saarlandes and Fraunhofer-Institut f\"ur Fertigungstechnik und Angewandte Materialforschung.

\section*{Conflicts of interest}
The authors declare no conflicts of interest.

\appendix
\section{Thermodynamic inequality}\label{thermo}
The principle virtual power satisfying the kinematic constraints is evaluated as the sum of internal and external contributions due to mechanical and micro force systems. The principle of virtual power takes the form:
\begin{equation}
	\dot{\cal{E}} = \cal{P}_{\rm mech} + \cal{P}_{\rm mic},
\end{equation}
where $\dot{\cal{E}}$, $\cal{P}_{\rm mech}$ and $\cal{P}_{\rm mic}$ are the total power, mechanical power and the power due to the micro force system. For  generalised virtual velocity ${\cal V} = \left\{\dot{\Tu}, \dot{\phi}\right\}$ the virtual power is
\begin{equation}
	\begin{aligned}
		 \int\limits_{\Omega}\TT\, :\, \grad\dot{\Tu}\,{\rm dV} + \int\limits_{\Omega} \omega \cdot \grad\dot{\phi}\,{\rm dV} + \int\limits_{\Omega} \varsigma \, \dot{\phi} \,{\rm dV} = \int\limits_{\partial\Omega}\Tt\cdot\dot{\Tu} \, {\rm dA} + \int\limits_{\Omega} \Tb \cdot \dot{\Tu}\, {\rm dV} + \int\limits_{\partial\Omega} \chi \, \dot{\phi} \, {\rm dA} + \int\limits_{\Omega}\Upsilon\, \dot{\phi} \, {\rm dV}, 
		\label{virtualpower}
	\end{aligned}
\end{equation}
where $\TT$ is the Cauchy stress tensor, $\Tt = \TT\cdot\Tn$ is traction, and $\Tb$ is the external body force. $\boldsymbol{\omega}$, $\varsigma$, $\chi$ and $\Upsilon$ are microscopic stress, microscopic internal force, external traction and external microscopic force of the micro force system \cite{GURTIN1996178}. Considering the virtual velocity ${\cal V} = \left(\dot{\Tu}, 0\right)$ and ${\cal V} = \left(0, \dot{\phi}\right)$ results in local forms of balance equations after applying divergence theorem
\begin{subequations}
	\begin{align}
		\div\, \TT + \Tb &= \mathbf{0} \label{bem}\\
		\div \,\boldsymbol{\omega} - \varsigma + \Upsilon &= 0.\label{bpfd}
	\end{align}
\end{subequations}
Equation \eqref{bem} and \eqref{bpfd} are the balance equations for momentum and phase-field equation for $\chi = \boldsymbol{\omega}\cdot\Tn$.
\subsection{Dissipation inequality}
The entropy inequality (Clausius-Duhem inequality) is required to formulate a thermodynamically consistent material law. The details of the thermodynamic evaluation of phase-field damage are complex due to the many terms involved. However, constitutive equations need to postulate to a material model and build relations between the kinematics and the balance equations. By assuming the Clausius-Duhem inequality as a condition for a non-negative entropy, the following inequality is obtained for an isothermal condition based on the micro force system \cite{FRIED1993326,Fried,GURTIN1996178}
\begin{equation}
	\TT:\TD + \omega\cdot\grad\dot{\phi} + \varsigma\, \dot{\phi} - \dot{W} \ge 0,
	\label{inequalit}
\end{equation}
where $\dot{W}$ denotes the rate of the free energy function. The free energy density of the phase-field damage model is given by the sum of the mechanical and the regularised fracture energy 
\begin{equation}
	\begin{aligned}
		W\left(J, \bar{\rm I}_1^{\bar{ \mathbf{B}}},\bar{\rm I}_1^{\bar{\mathbf{B}}^j_e} ,\phi,\grad\phi\right) = W_b\left(J, \bar{\rm I}_1^{\bar{ \mathbf{B}}},\bar{\rm I}_1^{\bar{\mathbf{B}}^j_e} ,\phi\right) \hspace{1mm}+ W_\phi (\phi,\grad\, \phi),
	\end{aligned}
	\label{full_ener}
\end{equation}
where the mechanical free energy function $W_b$ corresponds to the nearly incompressible viscoelasticity under large deformations. The free energy function of the finite-strain viscoelasticity is formulated as the algebraic sum of the volume-changing part $W_{\rm vol}$ and the isochoric parts consisting of the equilibrium $W_{\rm eq}$ and the non-equilibrium $W^j_{\rm neq}$ components \cite{HARTMANN20021439}. The non-equilibrium represents the overstresses due to rate-dependent properties represented by $j=1,2,...,n$ Maxwell elements. The mechanical free energy is a function of the damage variable and is formulated as 
\begin{equation}
	W_{\rm mech}\left(J, \bar{\rm I}_1^{\bar{\mathbf{B}}},\bar{\rm I}_1^{\bar{\mathbf{B}}^j_e} ,\phi\right) =  W_{\rm vol} \left(J,\phi \right) + W_{\rm eq} \left(\bar{\rm I}_1^{\bar{\mathbf{B}}},\phi \right) + \sum_{j=1}^{n} W^j_{\rm neq}\left(\bar{\rm I}_1^{\bar{\mathbf{B}}^j_e},\phi\right). \label{finvisko_450-1}
\end{equation}
As a result of inserting the equation \eqref{finvisko_450-1} in equation \eqref{full_ener}, the energy function is expressed as 
\begin{equation}
	\begin{aligned}
		W\left(J, \bar{\rm I}_1^{\bar{ \mathbf{B}}},\bar{\rm I}_1^{\bar{\mathbf{B}}^j_e} ,\phi,\grad\phi\right) = W_{\rm vol}(J,\phi) \hspace{4mm}+ W_{\rm eq}\left(\bar{\rm I}_1^{\bar{\mathbf{B}}},\phi \right)
		+ \sum_{j=1}^{n} W^j_{\rm neq}\left(\bar{\rm I}_1^{\bar{\mathbf{B}}^j_e},\phi \right) + \hspace{4mm} W_\phi (\phi,\grad\, \phi).
	\end{aligned}\label{free}
\end{equation}
The process variables to evaluate the inequality are \eqref{inequalit}
\begin{equation}
	\mathcal{S} = \left\{\mathbf{B},\mathbf{B}_e^j,\phi,\grad\, \phi \right\}.
\end{equation}
The time derivative of the free energy function $\dot{W}$ derived with the process variables  to express Clausius-Duhem inequality yields 
\begin{equation}
	\begin{aligned}
		\dot{W} =\,\, &\frac{\partial W_{\rm vol}\left(J, \phi\right)}{\partial \mathbf{B}} \colon \dot{ \mathbf{B}} + \frac{\partial W_{\rm eq} \left({\rm I}_1^{\bar{\mathbf{B}}},\phi \right)}{\partial \mathbf{B}} \colon \dot{ \mathbf{B}} + \sum_{j=1}^{n} \frac{\partial W^j_{\rm neq}\left({\rm I}_1^{\bar{\mathbf{B}}^j_e},\phi \right)}{\partial {\mathbf{B}}^j_e} \colon \dot{{\mathbf{B}}}^j_e  
		+\frac{\partial W}{\partial \phi} \colon \dot{\phi}+ \frac{\partial W}{\partial \grad \phi} \colon \grad\,\dot{\phi},
	\end{aligned}
	\label{free_der}
\end{equation}
and the equation \eqref{free_der} is applied to the inequality \eqref{inequalit} leading to the simplified form 
\begin{equation}
	\begin{aligned}
		&\left(-2\rho \mathbf{B}\cdot \frac{\partial W_{\rm vol}}{\partial \mathbf{B}} -2\rho{\mathbf{B}}\cdot\frac{\partial{W_{\rm eq}}}{\partial{\mathbf{B}}} - \sum_{j=1}^{n}2\rho\bar{\mathbf{B}}_e^j\cdot\frac{\partial{W_{\rm neq}}} {\partial\bar{\mathbf{B}}_e^j} + \TT\right):\mathbf{D}\,\\
		+\,
		&\hspace{1mm}\sum_{j=1}^{n}2\rho\frac{\partial{W_{\rm neq}}}{\partial\bar{\mathbf{B}}_e^j} : \left(\mathbf{F}_e^j\cdot\overset{\triangle}\TGamma {}_i^j\cdot \left(\mathbf{F}_e^j\right)^T \right) +\,\left(\varsigma-\frac{\partial W}{\partial \phi} \right)\cdot\dot{\phi}
		+\, \left(\omega-\frac{\partial W}{\partial\,\grad \phi} \right)\cdot\grad\,\dot{\phi} \geq 0.
	\end{aligned}
	\label{dissipation}
\end{equation}
where $\overset{\triangle}\TGamma {}_i^j$ is the inelastic deformation rate of the intermediate configuration. Based on the argumentation of Coleman \& Noll, the first term of the inequality leads to the constitutive equation for the stress tensor by introducing an assumption $W_{\left(\bullet\right)} = \rho W_{\left(\bullet\right)}$ \cite{HARTMANN20021439} for the free energy of the volumetric part and the isochoric parts of the equilibrium and the non-equilibrium parts
\begin{equation}
	\begin{aligned}
		\TT =\,2 \mathbf{B}\cdot \frac{\partial W_{\rm vol}}{\partial \mathbf{B}} +2{\mathbf{B}}\cdot\frac{\partial{W_{\rm eq}}}{\partial{\mathbf{B}}} + \sum_{j=1}^{n}2\bar{\mathbf{B}}_e^j\cdot\frac{\partial{W_{\rm neq}}} {\partial\bar{\mathbf{B}}_e^j}.
		\label{ratefreener}
	\end{aligned}
\end{equation}
The remaining inequality function leads to the residual dissipation equations concerning the evolution equations for the inelastic deformation rates of $n$ Maxwell elements \cite{Seldan,Lionphd}
\begin{equation}
	\begin{aligned}
		\sum_{j=1}^{n}2\rho\frac{\partial{W_{\rm neq}}}{\partial\bar{\mathbf{B}}_e^j} : \left(\mathbf{F}_e^j\cdot\overset{\triangle}\TGamma {}_i^j\cdot \left(\mathbf{F}_e^j\right)^T \right) \geq 0,
	\end{aligned}\label{simplediss}
\end{equation}
After some mathematical calculations and using the kinematic relations of finite deformation discussed earlier over the equation \eqref{simplediss} leads to the evolution equation\cite{Hauptpeter,Lionphd,Seldan}
\begin{equation}
	\begin{aligned}
		\dot{\bar{\TC}}_i^j\,&=\,\frac{4}{r_j} \left[\bar{\TC} - \frac{1}{3}{\rm{tr}}\left(\bar{\TC}\cdot\left(\bar{\TC}_i^j\right)^{-1} \right)\bar{\TC}_i^j\right],\\
	\end{aligned}
	\label{inelasticright}
\end{equation} 
further applying the inequality condition over the group of terms leads to 
\begin{equation}
	\begin{aligned}
		\left(\varsigma-\frac{\partial W}{\partial \phi} \right)\cdot\dot{\phi}
		+\,\hspace{1mm} \left(\omega-\frac{\partial W}{\partial\, \grad\phi} \right)\cdot\grad\,\dot{\phi} \geq 0
	\end{aligned}
	\label{dissipation1}
\end{equation}
and the consequent constitutive equations of the microscopic phase-field equation follows 
\begin{equation}
	\begin{aligned}
		\omega=\frac{\partial W_{\phi}(\phi,\grad\, \phi)}{\partial \phi},
		\hspace{10mm} \varsigma=\frac{\partial W_{\phi}(\phi,\grad\, \phi)}{\partial\, \grad\phi}. 
	\end{aligned}
	\label{dissipation2}
\end{equation}
Finally, the constitutive equations \eqref{dissipation2} are substituted in the equation \eqref{bpfd} to obtain the phase-field equation 
\begin{equation}
	\div\left(\frac{\partial W_{\phi}(\phi,\grad\, \phi)}{\partial\, \grad\phi} \right) - \frac{\partial W_{\phi}(\phi,\grad\, \phi)}{\partial \phi} = 0.
\end{equation}
Based on the micro force balance law, Gurtin \cite{GURTIN1996178} proposed the general form of evolution for the damage order parameter $\phi$ consistent with the equation \eqref{dissipation2} takes the form 
\begin{equation}
	\dot{\phi} = -M\left(W-E_c\left(\ell_f\,\div\phi-\frac{1}{\ell_f}\left(1-\phi\right)\right)\right),
\end{equation}
where $M>0$ is a scalar mobility parameter.

%
%
	
	\bibliographystyle{acm}
	\bibliography{Phase-field_damage}
	
\end{document}

%% file: defs.tex
\usepackage{amsfonts}
\usepackage{amssymb}
\usepackage{amsbsy}
\usepackage{ifthen}
\usepackage{multirow}
\usepackage{rotating}
\usepackage{color}
\usepackage{todonotes}
\usepackage{psfrag}
\usepackage{subcaption}
\usepackage{graphicx}

\newcommand{\beq}{\begin{equation}}

\newcommand{\eeq}[1]{%
  \ifthenelse{\equal{#1}{ }}
  {
    \end{equation}
  }
  {
    \label{eq:#1}
    \end{equation}
  }
}

\newcommand{\beqn}      {\begin{eqnarray}}
\newcommand{\eeqn}      {\end{eqnarray}}
\newcommand     {\bea}          {\begin{equationarray}}
\newcommand     {\eea}          {\end{equationarray}}
\newcommand{\beqa}{\begin{eqnarray}}
\newcommand{\eeqa}{\end{eqnarray}}

\newcommand{\bit}       {\begin{itemize}}
\newcommand{\eit}       {\end{itemize}}

\newcommand{\bde}       {\begin{description}}
\newcommand{\ede}       {\end{description}}

\newcommand{\bec}       {\begin{center}}
\newcommand{\eec}       {\end{center}}

\newsavebox{\BildText}
                        {\centerline{\usebox{\BildText}}\end{figure}}
                        
\newcommand{\Kasten}[3]%
{\begin{equation}
  \fbox{
    \begin{minipage}{5.5in}
      \centering
      #2
      \rule{5.4in}{0.5pt}
      {\bf #3}
    \end{minipage}
  }
  \label{box:#1}
\end{equation}}

\newcommand{\RefBraces}[1]%
{%
  \typeout{Referenz #1 in Seite \thepage}%
  \ifthenelse{\equal{\pageref{#1}}{\thepage}}%
             {(\ref{#1})}%
             {(\ref{#1}) auf Seite \pageref{#1}}%
}

\DeclareMathAlphabet\eufm{U}{euf}{m}{n}
\DeclareMathAlphabet\eufb{U}{euf}{b}{n}
\DeclareMathAlphabet\eurm{U}{eur}{m}{n}
\DeclareMathAlphabet\eurb{U}{eur}{b}{n}
\DeclareMathAlphabet\eusm{U}{eus}{m}{n}
\DeclareMathAlphabet\eusb{U}{eus}{b}{n}
\DeclareMathAlphabet\mathsfm{OT1}{cmss}{m}{n}
\DeclareMathAlphabet\mathsfb{OT1}{cmss}{bx}{n}

\newcommand{\Bs}[1]{\boldsymbol{#1}}            
\newcommand{\Qed}{\hfill \mbox{$\Box$}}         

\renewcommand{\div}    {\hbox{\rmfamily div}}               

\newcommand{\bigo}[1]{\setbox0\hbox{$\bigcirc$}
             \rlap{\raise .2ex\hbox to \wd0{\hfil ${\scriptscriptstyle
                   #1}$\hfil}}\box0}
 
\newcommand{\TenB}[1]{{\rm\bf #1}}              
\newcommand{\TenG}[1]{\Bs{#1}}                  
\newcommand{\TenF}[1]{\stackrel{4}{{\eurb #1}}} 

\newcommand{\TB}        {\TenB{B}}
\newcommand{\TC}        {\TenB{C}}
\newcommand{\TD}        {\TenB{D}}

\newcommand{\TF}        {\TenB{F}}

\newcommand{\TS}        {\TenB{S}}
\newcommand{\TT}        {\TenB{T}}

\newcommand{\Tb}        {\TenB{b}}

\newcommand{\Tn}        {\TenB{n}}

\newcommand{\Tt}        {\TenB{t}}

\newcommand{\Tu}        {\TenB{u}}

\newcommand{\Tx}        {\TenB{x}}

\newcommand{\kapp}        {\boldsymbol{\kappa}}

\newcommand{\TGamma}    {\TenG{\Gamma}}

\newcommand{\svektor}[2]%
{\left(\!\!\!\begin{array}{c} #1\\#2 \end{array}\!\!\!\right)}

\newcommand{\zvektor}[2]%
{\left(#1,#2\right)^T}

\newcommand{\Svektor}[2]%
{\left[\!\!\!\begin{array}{l} #1\\#2 \end{array}\!\!\!\right]}

\newcommand{\Zvektor}[2]%
{\left[#1,#2\right]^T}

%% file: MatModel/rheologisches_modell1.eps_tex
\begingroup%
  \makeatletter%
  \providecommand\color[2][]{%
    \errmessage{(Inkscape) Color is used for the text in Inkscape, but the package 'color.sty' is not loaded}%
    \renewcommand\color[2][]{}%
  }%
  \providecommand\transparent[1]{%
    \errmessage{(Inkscape) Transparency is used (non-zero) for the text in Inkscape, but the package 'transparent.sty' is not loaded}%
    \renewcommand\transparent[1]{}%
  }%
  \providecommand\rotatebox[2]{#2}%
  \ifx\svgwidth\undefined%
    \setlength{\unitlength}{1376.00006104bp}%
    \ifx\svgscale\undefined%
      \relax%
    \else%
      \setlength{\unitlength}{\unitlength * \real{\svgscale}}%
    \fi%
  \else%
    \setlength{\unitlength}{\svgwidth}%
  \fi%
  \global\let\svgwidth\undefined%
  \global\let\svgscale\undefined%
  \makeatother%
  \begin{picture}(1,0.42877906)%
    \put(0,0){\includegraphics[width=\unitlength]{MatModel/rheologisches_modell1.eps}}%
    \put(-0.05550893,0.20363679){\color[rgb]{0,0,0}\makebox(0,0)[lb]{\smash{$\mu_{1}$}}}%
    \put(0.16443296,0.27494563){\color[rgb]{0,0,0}\makebox(0,0)[lb]{\smash{$\mu_{11}$}}}%
    \put(0.422265,0.27494563){\color[rgb]{0,0,0}\makebox(0,0)[lb]{\smash{$\mu_{12}$}}}%
    \put(0.77456092,0.27494563){\color[rgb]{0,0,0}\makebox(0,0)[lb]{\smash{$\mu_{1n}$}}}%
    \put(0.17642424,0.1059776){\color[rgb]{0,0,0}\makebox(0,0)[lb]{\smash{$\eta_1$}}}%
    \put(0.44193518,0.1059776){\color[rgb]{0,0,0}\makebox(0,0)[lb]{\smash{$\eta_2$}}}%
    \put(0.78804765,0.1059776){\color[rgb]{0,0,0}\makebox(0,0)[lb]{\smash{$\eta_n$}}}%
  \end{picture}%
\endgroup%

%% file: phasefield/angular_spec.eps_tex
\begingroup%
\vspace{-1cm}
  \makeatletter%
  \providecommand\color[2][]{%
    \errmessage{(Inkscape) Color is used for the text in Inkscape, but the package 'color.sty' is not loaded}%
    \renewcommand\color[2][]{}%
  }%
  \providecommand\transparent[1]{%
    \errmessage{(Inkscape) Transparency is used (non-zero) for the text in Inkscape, but the package 'transparent.sty' is not loaded}%
    \renewcommand\transparent[1]{}%
  }%
  \providecommand\rotatebox[2]{#2}%
  \newcommand*\fsize{\dimexpr\f@size pt\relax}%
  \newcommand*\lineheight[1]{\fontsize{\fsize}{#1\fsize}\selectfont}%
  \ifx\svgwidth\undefined%
    \setlength{\unitlength}{657.63779528bp}%
    \ifx\svgscale\undefined%
      \relax%
    \else%
      \setlength{\unitlength}{\unitlength * \real{\svgscale}}%
    \fi%
  \else%
    \setlength{\unitlength}{\svgwidth}%
  \fi%
  \global\let\svgwidth\undefined%
  \global\let\svgscale\undefined%
  \makeatother%
  \begin{picture}(1,0.67672414)%
    \lineheight{1}%
    \setlength\tabcolsep{0pt}%
    \put(0,0){\includegraphics[width=\unitlength]{phasefield/angular_spec.eps}}%
    \put(0.41098363,0.58607728){\color[rgb]{1,1,1}\makebox(0,0)[lt]{\lineheight{1.25}\smash{\begin{tabular}[t]{l}$100$\end{tabular}}}}%
    \put(0.42,0.45908071){\color[rgb]{0,0,0}\makebox(0,0)[lt]{\lineheight{1.25}\smash{\begin{tabular}[t]{l}$90^\circ \pm 0.5^\circ$\end{tabular}}}}%
    \put(0.78,0.33){\color[rgb]{0,0,0}\makebox(0,0)[lt]{\lineheight{1.25}\smash{\begin{tabular}[t]{l}$\approx 22$\end{tabular}}}}%
    \put(0.22,0.28){\color[rgb]{0,0,0}\rotatebox{63.77423155}{\makebox(0,0)[lt]{\lineheight{1.25}\smash{\begin{tabular}[t]{l}${\rm R}19 \pm 0.05$\end{tabular}}}}}%
    \put(0.96,0.29){\color[rgb]{0,0,0}\rotatebox{88.54781331}{\makebox(0,0)[lt]{\lineheight{1.25}\smash{\begin{tabular}[t]{l}$19 \pm 0.05$\end{tabular}}}}}%
    \put(0.26,0.06){\color[rgb]{0,0,0}\rotatebox{47.20994265}{\makebox(0,0)[lt]{\lineheight{1.25}\smash{\begin{tabular}[t]{l}${\rm R}12.7 \pm 0.05$\end{tabular}}}}}%
    \put(0.63,0.21){\color[rgb]{0,0,0}\rotatebox{-55.50078137}{\makebox(0,0)[lt]{\lineheight{1.25}\smash{\begin{tabular}[t]{l}${\rm R}25.4 \pm 0.05$\end{tabular}}}}}%
  \end{picture}%
\endgroup%

%% file: phasefield/crack2VST.tex
\begingroup
\large
  \makeatletter
  \providecommand\color[2][]{%
    \GenericError{(gnuplot) \space\space\space\@spaces}{%
      Package color not loaded in conjunction with
      terminal option `colourtext'%
    }{See the gnuplot documentation for explanation.%
    }{Either use 'blacktext' in gnuplot or load the package
      color.sty in LaTeX.}%
    \renewcommand\color[2][]{}%
  }%
  \providecommand\includegraphics[2][]{%
    \GenericError{(gnuplot) \space\space\space\@spaces}{%
      Package graphicx or graphics not loaded%
    }{See the gnuplot documentation for explanation.%
    }{The gnuplot epslatex terminal needs graphicx.sty or graphics.sty.}%
    \renewcommand\includegraphics[2][]{}%
  }%
  \providecommand\rotatebox[2]{#2}%
  \@ifundefined{ifGPcolor}{%
    \newif\ifGPcolor
    \GPcolortrue
  }{}%
  \@ifundefined{ifGPblacktext}{%
    \newif\ifGPblacktext
    \GPblacktextfalse
  }{}%
  \let\gplgaddtomacro\g@addto@macro
  \gdef\gplbacktext{}%
  \gdef\gplfronttext{}%
  \makeatother
  \ifGPblacktext
    \def\colorrgb#1{}%
    \def\colorgray#1{}%
  \else
    \ifGPcolor
      \def\colorrgb#1{\color[rgb]{#1}}%
      \def\colorgray#1{\color[gray]{#1}}%
      \expandafter\def\csname LTw\endcsname{\color{white}}%
      \expandafter\def\csname LTb\endcsname{\color{black}}%
      \expandafter\def\csname LTa\endcsname{\color{black}}%
      \expandafter\def\csname LT0\endcsname{\color[rgb]{1,0,0}}%
      \expandafter\def\csname LT1\endcsname{\color[rgb]{0,1,0}}%
      \expandafter\def\csname LT2\endcsname{\color[rgb]{0,0,1}}%
      \expandafter\def\csname LT3\endcsname{\color[rgb]{1,0,1}}%
      \expandafter\def\csname LT4\endcsname{\color[rgb]{0,1,1}}%
      \expandafter\def\csname LT5\endcsname{\color[rgb]{1,1,0}}%
      \expandafter\def\csname LT6\endcsname{\color[rgb]{0,0,0}}%
      \expandafter\def\csname LT7\endcsname{\color[rgb]{1,0.3,0}}%
      \expandafter\def\csname LT8\endcsname{\color[rgb]{0.5,0.5,0.5}}%
    \else
      \def\colorrgb#1{\color{black}}%
      \def\colorgray#1{\color[gray]{#1}}%
      \expandafter\def\csname LTw\endcsname{\color{white}}%
      \expandafter\def\csname LTb\endcsname{\color{black}}%
      \expandafter\def\csname LTa\endcsname{\color{black}}%
      \expandafter\def\csname LT0\endcsname{\color{black}}%
      \expandafter\def\csname LT1\endcsname{\color{black}}%
      \expandafter\def\csname LT2\endcsname{\color{black}}%
      \expandafter\def\csname LT3\endcsname{\color{black}}%
      \expandafter\def\csname LT4\endcsname{\color{black}}%
      \expandafter\def\csname LT5\endcsname{\color{black}}%
      \expandafter\def\csname LT6\endcsname{\color{black}}%
      \expandafter\def\csname LT7\endcsname{\color{black}}%
      \expandafter\def\csname LT8\endcsname{\color{black}}%
    \fi
  \fi
  \setlength{\unitlength}{0.0500bp}%
  \begin{picture}(7200.00,5040.00)%
    \gplgaddtomacro\gplbacktext{%
      \csname LTb\endcsname%
      \put(682,704){\makebox(0,0)[r]{\strut{} 0}}%
      \put(682,1518){\makebox(0,0)[r]{\strut{} 5}}%
      \put(682,2332){\makebox(0,0)[r]{\strut{} 10}}%
      \put(682,3147){\makebox(0,0)[r]{\strut{} 15}}%
      \put(682,3961){\makebox(0,0)[r]{\strut{} 20}}%
      \put(682,4775){\makebox(0,0)[r]{\strut{} 25}}%
      \put(814,484){\makebox(0,0){\strut{} 0}}%
      \put(1563,484){\makebox(0,0){\strut{} 1}}%
      \put(2311,484){\makebox(0,0){\strut{} 2}}%
      \put(3060,484){\makebox(0,0){\strut{} 3}}%
      \put(3809,484){\makebox(0,0){\strut{} 4}}%
      \put(4557,484){\makebox(0,0){\strut{} 5}}%
      \put(5306,484){\makebox(0,0){\strut{} 6}}%
      \put(6054,484){\makebox(0,0){\strut{} 7}}%
      \put(6803,484){\makebox(0,0){\strut{} 8}}%
      \put(176,2739){\rotatebox{-270}{\makebox(0,0){\strut{}Force $\left[\rm N\right]$}}}%
      \put(4072,154){\makebox(0,0){\strut{}Displacement $\left[\rm mm\right]$}}%
    }%
    \gplgaddtomacro\gplfronttext{%
      \csname LTb\endcsname%
      \put(3322,4514){\makebox(0,0)[r]{\strut{}Standard deviation}}%
      \csname LTb\endcsname%
      \put(3322,4118){\makebox(0,0)[r]{\strut{}Experiment}}%
      \csname LTb\endcsname%
      \put(3322,3722){\makebox(0,0)[r]{\strut{}Simulation}}%
    }%
    \gplbacktext
    \put(0,0){\includegraphics{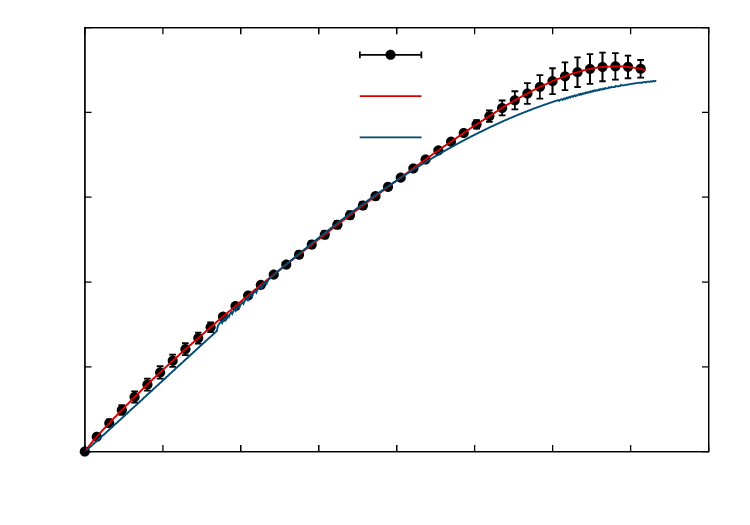}}%
    \gplfronttext
  \end{picture}%
\endgroup

%% file: phasefield/29RH_KVSD.tex
\begingroup
\large
  \makeatletter
  \providecommand\color[2][]{%
    \GenericError{(gnuplot) \space\space\space\@spaces}{%
      Package color not loaded in conjunction with
      terminal option `colourtext'%
    }{See the gnuplot documentation for explanation.%
    }{Either use 'blacktext' in gnuplot or load the package
      color.sty in LaTeX.}%
    \renewcommand\color[2][]{}%
  }%
  \providecommand\includegraphics[2][]{%
    \GenericError{(gnuplot) \space\space\space\@spaces}{%
      Package graphicx or graphics not loaded%
    }{See the gnuplot documentation for explanation.%
    }{The gnuplot epslatex terminal needs graphicx.sty or graphics.sty.}%
    \renewcommand\includegraphics[2][]{}%
  }%
  \providecommand\rotatebox[2]{#2}%
  \@ifundefined{ifGPcolor}{%
    \newif\ifGPcolor
    \GPcolortrue
  }{}%
  \@ifundefined{ifGPblacktext}{%
    \newif\ifGPblacktext
    \GPblacktextfalse
  }{}%
  \let\gplgaddtomacro\g@addto@macro
  \gdef\gplbacktext{}%
  \gdef\gplfronttext{}%
  \makeatother
  \ifGPblacktext
    \def\colorrgb#1{}%
    \def\colorgray#1{}%
  \else
    \ifGPcolor
      \def\colorrgb#1{\color[rgb]{#1}}%
      \def\colorgray#1{\color[gray]{#1}}%
      \expandafter\def\csname LTw\endcsname{\color{white}}%
      \expandafter\def\csname LTb\endcsname{\color{black}}%
      \expandafter\def\csname LTa\endcsname{\color{black}}%
      \expandafter\def\csname LT0\endcsname{\color[rgb]{1,0,0}}%
      \expandafter\def\csname LT1\endcsname{\color[rgb]{0,1,0}}%
      \expandafter\def\csname LT2\endcsname{\color[rgb]{0,0,1}}%
      \expandafter\def\csname LT3\endcsname{\color[rgb]{1,0,1}}%
      \expandafter\def\csname LT4\endcsname{\color[rgb]{0,1,1}}%
      \expandafter\def\csname LT5\endcsname{\color[rgb]{1,1,0}}%
      \expandafter\def\csname LT6\endcsname{\color[rgb]{0,0,0}}%
      \expandafter\def\csname LT7\endcsname{\color[rgb]{1,0.3,0}}%
      \expandafter\def\csname LT8\endcsname{\color[rgb]{0.5,0.5,0.5}}%
    \else
      \def\colorrgb#1{\color{black}}%
      \def\colorgray#1{\color[gray]{#1}}%
      \expandafter\def\csname LTw\endcsname{\color{white}}%
      \expandafter\def\csname LTb\endcsname{\color{black}}%
      \expandafter\def\csname LTa\endcsname{\color{black}}%
      \expandafter\def\csname LT0\endcsname{\color{black}}%
      \expandafter\def\csname LT1\endcsname{\color{black}}%
      \expandafter\def\csname LT2\endcsname{\color{black}}%
      \expandafter\def\csname LT3\endcsname{\color{black}}%
      \expandafter\def\csname LT4\endcsname{\color{black}}%
      \expandafter\def\csname LT5\endcsname{\color{black}}%
      \expandafter\def\csname LT6\endcsname{\color{black}}%
      \expandafter\def\csname LT7\endcsname{\color{black}}%
      \expandafter\def\csname LT8\endcsname{\color{black}}%
    \fi
  \fi
  \setlength{\unitlength}{0.0500bp}%
  \begin{picture}(7200.00,5040.00)%
    \gplgaddtomacro\gplbacktext{%
      \csname LTb\endcsname%
      \put(682,704){\makebox(0,0)[r]{\strut{} 0}}%
      \put(682,1518){\makebox(0,0)[r]{\strut{} 5}}%
      \put(682,2332){\makebox(0,0)[r]{\strut{} 10}}%
      \put(682,3147){\makebox(0,0)[r]{\strut{} 15}}%
      \put(682,3961){\makebox(0,0)[r]{\strut{} 20}}%
      \put(682,4775){\makebox(0,0)[r]{\strut{} 25}}%
      \put(814,484){\makebox(0,0){\strut{} 0}}%
      \put(1479,484){\makebox(0,0){\strut{} 1}}%
      \put(2145,484){\makebox(0,0){\strut{} 2}}%
      \put(2810,484){\makebox(0,0){\strut{} 3}}%
      \put(3476,484){\makebox(0,0){\strut{} 4}}%
      \put(4141,484){\makebox(0,0){\strut{} 5}}%
      \put(4807,484){\makebox(0,0){\strut{} 6}}%
      \put(5472,484){\makebox(0,0){\strut{} 7}}%
      \put(6138,484){\makebox(0,0){\strut{} 8}}%
      \put(6803,484){\makebox(0,0){\strut{} 9}}%
      \put(176,2739){\rotatebox{-270}{\makebox(0,0){\strut{}Force $\left[\rm N\right]$}}}%
      \put(4072,154){\makebox(0,0){\strut{}Displacement $\left[\rm mm\right]$}}%
    }%
    \gplgaddtomacro\gplfronttext{%
      \csname LTb\endcsname%
      \put(3586,4514){\makebox(0,0)[r]{\strut{}Standard deviation}}%
      \csname LTb\endcsname%
      \put(3586,4118){\makebox(0,0)[r]{\strut{}29$\%$ r.H. Experiment}}%
      \csname LTb\endcsname%
      \put(3586,3722){\makebox(0,0)[r]{\strut{}29$\%$ r.H. Simulation}}%
    }%
    \gplgaddtomacro\gplbacktext{%
      \csname LTb\endcsname%
      \put(682,704){\makebox(0,0)[r]{\strut{} 0}}%
      \put(682,1518){\makebox(0,0)[r]{\strut{} 5}}%
      \put(682,2332){\makebox(0,0)[r]{\strut{} 10}}%
      \put(682,3147){\makebox(0,0)[r]{\strut{} 15}}%
      \put(682,3961){\makebox(0,0)[r]{\strut{} 20}}%
      \put(682,4775){\makebox(0,0)[r]{\strut{} 25}}%
      \put(814,484){\makebox(0,0){\strut{} 0}}%
      \put(1479,484){\makebox(0,0){\strut{} 1}}%
      \put(2145,484){\makebox(0,0){\strut{} 2}}%
      \put(2810,484){\makebox(0,0){\strut{} 3}}%
      \put(3476,484){\makebox(0,0){\strut{} 4}}%
      \put(4141,484){\makebox(0,0){\strut{} 5}}%
      \put(4807,484){\makebox(0,0){\strut{} 6}}%
      \put(5472,484){\makebox(0,0){\strut{} 7}}%
      \put(6138,484){\makebox(0,0){\strut{} 8}}%
      \put(6803,484){\makebox(0,0){\strut{} 9}}%
      \put(176,2739){\rotatebox{-270}{\makebox(0,0){\strut{}Force $\left[\rm N\right]$}}}%
      \put(4072,154){\makebox(0,0){\strut{}Displacement $\left[\rm mm\right]$}}%
    }%
    \gplgaddtomacro\gplfronttext{%
      \csname LTb\endcsname%
      \put(3586,4514){\makebox(0,0)[r]{\strut{}Standard deviation}}%
      \csname LTb\endcsname%
      \put(3586,4118){\makebox(0,0)[r]{\strut{}29$\%$ r.H. Experiment}}%
      \csname LTb\endcsname%
      \put(3586,3722){\makebox(0,0)[r]{\strut{}29$\%$ r.H. Simulation}}%
    }%
    \gplbacktext
    \put(0,0){\includegraphics{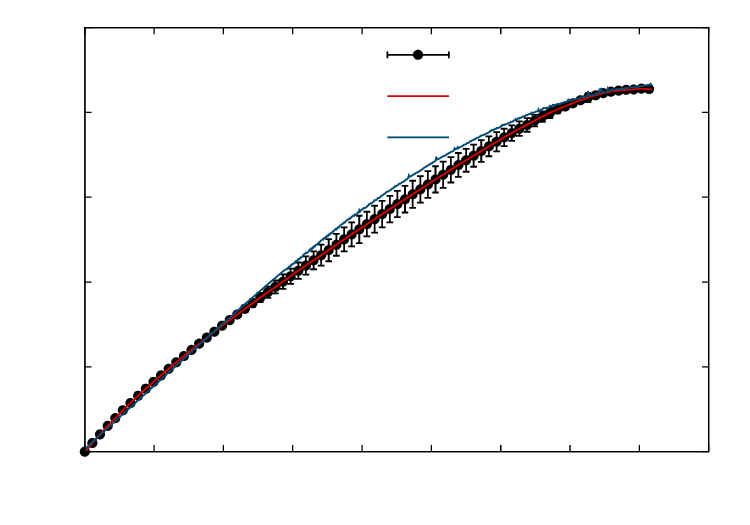}}%
    \gplfronttext
  \end{picture}%
\endgroup

%% file: phasefield/67RH_KVSD.tex
\begingroup
\large
  \makeatletter
  \providecommand\color[2][]{%
    \GenericError{(gnuplot) \space\space\space\@spaces}{%
      Package color not loaded in conjunction with
      terminal option `colourtext'%
    }{See the gnuplot documentation for explanation.%
    }{Either use 'blacktext' in gnuplot or load the package
      color.sty in LaTeX.}%
    \renewcommand\color[2][]{}%
  }%
  \providecommand\includegraphics[2][]{%
    \GenericError{(gnuplot) \space\space\space\@spaces}{%
      Package graphicx or graphics not loaded%
    }{See the gnuplot documentation for explanation.%
    }{The gnuplot epslatex terminal needs graphicx.sty or graphics.sty.}%
    \renewcommand\includegraphics[2][]{}%
  }%
  \providecommand\rotatebox[2]{#2}%
  \@ifundefined{ifGPcolor}{%
    \newif\ifGPcolor
    \GPcolortrue
  }{}%
  \@ifundefined{ifGPblacktext}{%
    \newif\ifGPblacktext
    \GPblacktextfalse
  }{}%
  \let\gplgaddtomacro\g@addto@macro
  \gdef\gplbacktext{}%
  \gdef\gplfronttext{}%
  \makeatother
  \ifGPblacktext
    \def\colorrgb#1{}%
    \def\colorgray#1{}%
  \else
    \ifGPcolor
      \def\colorrgb#1{\color[rgb]{#1}}%
      \def\colorgray#1{\color[gray]{#1}}%
      \expandafter\def\csname LTw\endcsname{\color{white}}%
      \expandafter\def\csname LTb\endcsname{\color{black}}%
      \expandafter\def\csname LTa\endcsname{\color{black}}%
      \expandafter\def\csname LT0\endcsname{\color[rgb]{1,0,0}}%
      \expandafter\def\csname LT1\endcsname{\color[rgb]{0,1,0}}%
      \expandafter\def\csname LT2\endcsname{\color[rgb]{0,0,1}}%
      \expandafter\def\csname LT3\endcsname{\color[rgb]{1,0,1}}%
      \expandafter\def\csname LT4\endcsname{\color[rgb]{0,1,1}}%
      \expandafter\def\csname LT5\endcsname{\color[rgb]{1,1,0}}%
      \expandafter\def\csname LT6\endcsname{\color[rgb]{0,0,0}}%
      \expandafter\def\csname LT7\endcsname{\color[rgb]{1,0.3,0}}%
      \expandafter\def\csname LT8\endcsname{\color[rgb]{0.5,0.5,0.5}}%
    \else
      \def\colorrgb#1{\color{black}}%
      \def\colorgray#1{\color[gray]{#1}}%
      \expandafter\def\csname LTw\endcsname{\color{white}}%
      \expandafter\def\csname LTb\endcsname{\color{black}}%
      \expandafter\def\csname LTa\endcsname{\color{black}}%
      \expandafter\def\csname LT0\endcsname{\color{black}}%
      \expandafter\def\csname LT1\endcsname{\color{black}}%
      \expandafter\def\csname LT2\endcsname{\color{black}}%
      \expandafter\def\csname LT3\endcsname{\color{black}}%
      \expandafter\def\csname LT4\endcsname{\color{black}}%
      \expandafter\def\csname LT5\endcsname{\color{black}}%
      \expandafter\def\csname LT6\endcsname{\color{black}}%
      \expandafter\def\csname LT7\endcsname{\color{black}}%
      \expandafter\def\csname LT8\endcsname{\color{black}}%
    \fi
  \fi
  \setlength{\unitlength}{0.0500bp}%
  \begin{picture}(7200.00,5040.00)%
    \gplgaddtomacro\gplbacktext{%
      \csname LTb\endcsname%
      \put(682,704){\makebox(0,0)[r]{\strut{} 0}}%
      \put(682,1518){\makebox(0,0)[r]{\strut{} 5}}%
      \put(682,2332){\makebox(0,0)[r]{\strut{} 10}}%
      \put(682,3147){\makebox(0,0)[r]{\strut{} 15}}%
      \put(682,3961){\makebox(0,0)[r]{\strut{} 20}}%
      \put(682,4775){\makebox(0,0)[r]{\strut{} 25}}%
      \put(814,484){\makebox(0,0){\strut{} 0}}%
      \put(1479,484){\makebox(0,0){\strut{} 1}}%
      \put(2145,484){\makebox(0,0){\strut{} 2}}%
      \put(2810,484){\makebox(0,0){\strut{} 3}}%
      \put(3476,484){\makebox(0,0){\strut{} 4}}%
      \put(4141,484){\makebox(0,0){\strut{} 5}}%
      \put(4807,484){\makebox(0,0){\strut{} 6}}%
      \put(5472,484){\makebox(0,0){\strut{} 7}}%
      \put(6138,484){\makebox(0,0){\strut{} 8}}%
      \put(6803,484){\makebox(0,0){\strut{} 9}}%
      \put(176,2739){\rotatebox{-270}{\makebox(0,0){\strut{}Force $\left[\rm N\right]$}}}%
      \put(4072,154){\makebox(0,0){\strut{}Displacement $\left[\rm mm\right]$}}%
    }%
    \gplgaddtomacro\gplfronttext{%
      \csname LTb\endcsname%
      \put(3586,4514){\makebox(0,0)[r]{\strut{}Standard deviation}}%
      \csname LTb\endcsname%
      \put(3586,4118){\makebox(0,0)[r]{\strut{}67$\%$ r.H. Experiment}}%
      \csname LTb\endcsname%
      \put(3586,3722){\makebox(0,0)[r]{\strut{}67$\%$ r.H. Simulation}}%
    }%
    \gplgaddtomacro\gplbacktext{%
      \csname LTb\endcsname%
      \put(682,704){\makebox(0,0)[r]{\strut{} 0}}%
      \put(682,1518){\makebox(0,0)[r]{\strut{} 5}}%
      \put(682,2332){\makebox(0,0)[r]{\strut{} 10}}%
      \put(682,3147){\makebox(0,0)[r]{\strut{} 15}}%
      \put(682,3961){\makebox(0,0)[r]{\strut{} 20}}%
      \put(682,4775){\makebox(0,0)[r]{\strut{} 25}}%
      \put(814,484){\makebox(0,0){\strut{} 0}}%
      \put(1479,484){\makebox(0,0){\strut{} 1}}%
      \put(2145,484){\makebox(0,0){\strut{} 2}}%
      \put(2810,484){\makebox(0,0){\strut{} 3}}%
      \put(3476,484){\makebox(0,0){\strut{} 4}}%
      \put(4141,484){\makebox(0,0){\strut{} 5}}%
      \put(4807,484){\makebox(0,0){\strut{} 6}}%
      \put(5472,484){\makebox(0,0){\strut{} 7}}%
      \put(6138,484){\makebox(0,0){\strut{} 8}}%
      \put(6803,484){\makebox(0,0){\strut{} 9}}%
      \put(176,2739){\rotatebox{-270}{\makebox(0,0){\strut{}Force $\left[\rm N\right]$}}}%
      \put(4072,154){\makebox(0,0){\strut{}Displacement $\left[\rm mm\right]$}}%
    }%
    \gplgaddtomacro\gplfronttext{%
      \csname LTb\endcsname%
      \put(3586,4514){\makebox(0,0)[r]{\strut{}Standard deviation}}%
      \csname LTb\endcsname%
      \put(3586,4118){\makebox(0,0)[r]{\strut{}67$\%$ r.H. Experiment}}%
      \csname LTb\endcsname%
      \put(3586,3722){\makebox(0,0)[r]{\strut{}67$\%$ r.H. Simulation}}%
    }%
    \gplbacktext
    \put(0,0){\includegraphics{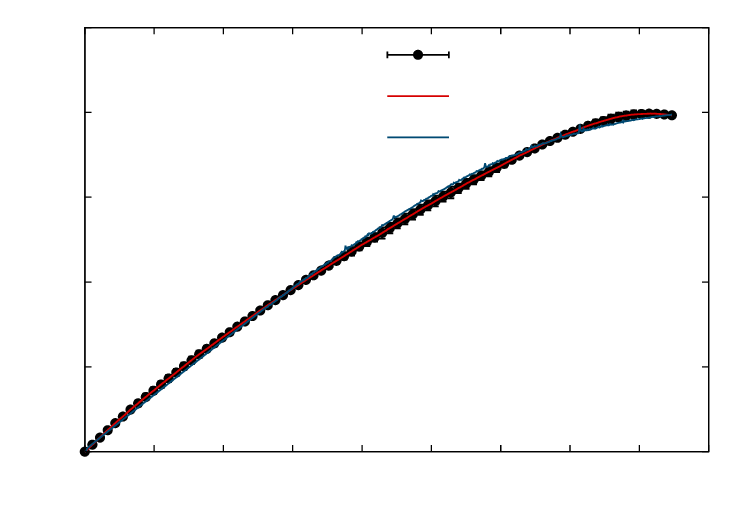}}%
    \gplfronttext
  \end{picture}%
\endgroup

%% file: phasefield/100RH_KVSD.tex
\begingroup
\large
  \makeatletter
  \providecommand\color[2][]{%
    \GenericError{(gnuplot) \space\space\space\@spaces}{%
      Package color not loaded in conjunction with
      terminal option `colourtext'%
    }{See the gnuplot documentation for explanation.%
    }{Either use 'blacktext' in gnuplot or load the package
      color.sty in LaTeX.}%
    \renewcommand\color[2][]{}%
  }%
  \providecommand\includegraphics[2][]{%
    \GenericError{(gnuplot) \space\space\space\@spaces}{%
      Package graphicx or graphics not loaded%
    }{See the gnuplot documentation for explanation.%
    }{The gnuplot epslatex terminal needs graphicx.sty or graphics.sty.}%
    \renewcommand\includegraphics[2][]{}%
  }%
  \providecommand\rotatebox[2]{#2}%
  \@ifundefined{ifGPcolor}{%
    \newif\ifGPcolor
    \GPcolortrue
  }{}%
  \@ifundefined{ifGPblacktext}{%
    \newif\ifGPblacktext
    \GPblacktextfalse
  }{}%
  \let\gplgaddtomacro\g@addto@macro
  \gdef\gplbacktext{}%
  \gdef\gplfronttext{}%
  \makeatother
  \ifGPblacktext
    \def\colorrgb#1{}%
    \def\colorgray#1{}%
  \else
    \ifGPcolor
      \def\colorrgb#1{\color[rgb]{#1}}%
      \def\colorgray#1{\color[gray]{#1}}%
      \expandafter\def\csname LTw\endcsname{\color{white}}%
      \expandafter\def\csname LTb\endcsname{\color{black}}%
      \expandafter\def\csname LTa\endcsname{\color{black}}%
      \expandafter\def\csname LT0\endcsname{\color[rgb]{1,0,0}}%
      \expandafter\def\csname LT1\endcsname{\color[rgb]{0,1,0}}%
      \expandafter\def\csname LT2\endcsname{\color[rgb]{0,0,1}}%
      \expandafter\def\csname LT3\endcsname{\color[rgb]{1,0,1}}%
      \expandafter\def\csname LT4\endcsname{\color[rgb]{0,1,1}}%
      \expandafter\def\csname LT5\endcsname{\color[rgb]{1,1,0}}%
      \expandafter\def\csname LT6\endcsname{\color[rgb]{0,0,0}}%
      \expandafter\def\csname LT7\endcsname{\color[rgb]{1,0.3,0}}%
      \expandafter\def\csname LT8\endcsname{\color[rgb]{0.5,0.5,0.5}}%
    \else
      \def\colorrgb#1{\color{black}}%
      \def\colorgray#1{\color[gray]{#1}}%
      \expandafter\def\csname LTw\endcsname{\color{white}}%
      \expandafter\def\csname LTb\endcsname{\color{black}}%
      \expandafter\def\csname LTa\endcsname{\color{black}}%
      \expandafter\def\csname LT0\endcsname{\color{black}}%
      \expandafter\def\csname LT1\endcsname{\color{black}}%
      \expandafter\def\csname LT2\endcsname{\color{black}}%
      \expandafter\def\csname LT3\endcsname{\color{black}}%
      \expandafter\def\csname LT4\endcsname{\color{black}}%
      \expandafter\def\csname LT5\endcsname{\color{black}}%
      \expandafter\def\csname LT6\endcsname{\color{black}}%
      \expandafter\def\csname LT7\endcsname{\color{black}}%
      \expandafter\def\csname LT8\endcsname{\color{black}}%
    \fi
  \fi
  \setlength{\unitlength}{0.0500bp}%
  \begin{picture}(7200.00,5040.00)%
    \gplgaddtomacro\gplbacktext{%
      \csname LTb\endcsname%
      \put(682,704){\makebox(0,0)[r]{\strut{} 0}}%
      \put(682,1518){\makebox(0,0)[r]{\strut{} 5}}%
      \put(682,2332){\makebox(0,0)[r]{\strut{} 10}}%
      \put(682,3147){\makebox(0,0)[r]{\strut{} 15}}%
      \put(682,3961){\makebox(0,0)[r]{\strut{} 20}}%
      \put(682,4775){\makebox(0,0)[r]{\strut{} 25}}%
      \put(814,484){\makebox(0,0){\strut{} 0}}%
      \put(1479,484){\makebox(0,0){\strut{} 1}}%
      \put(2145,484){\makebox(0,0){\strut{} 2}}%
      \put(2810,484){\makebox(0,0){\strut{} 3}}%
      \put(3476,484){\makebox(0,0){\strut{} 4}}%
      \put(4141,484){\makebox(0,0){\strut{} 5}}%
      \put(4807,484){\makebox(0,0){\strut{} 6}}%
      \put(5472,484){\makebox(0,0){\strut{} 7}}%
      \put(6138,484){\makebox(0,0){\strut{} 8}}%
      \put(6803,484){\makebox(0,0){\strut{} 9}}%
      \put(176,2739){\rotatebox{-270}{\makebox(0,0){\strut{}Force $\left[\rm N\right]$}}}%
      \put(4072,154){\makebox(0,0){\strut{}Displacement $\left[\rm mm\right]$}}%
    }%
    \gplgaddtomacro\gplfronttext{%
      \csname LTb\endcsname%
      \put(3718,4514){\makebox(0,0)[r]{\strut{}Standard deviation}}%
      \csname LTb\endcsname%
      \put(3718,4118){\makebox(0,0)[r]{\strut{}100$\%$ r.H. Experiment}}%
      \csname LTb\endcsname%
      \put(3718,3722){\makebox(0,0)[r]{\strut{}100$\%$ r.H. Simulation}}%
    }%
    \gplgaddtomacro\gplbacktext{%
      \csname LTb\endcsname%
      \put(682,704){\makebox(0,0)[r]{\strut{} 0}}%
      \put(682,1518){\makebox(0,0)[r]{\strut{} 5}}%
      \put(682,2332){\makebox(0,0)[r]{\strut{} 10}}%
      \put(682,3147){\makebox(0,0)[r]{\strut{} 15}}%
      \put(682,3961){\makebox(0,0)[r]{\strut{} 20}}%
      \put(682,4775){\makebox(0,0)[r]{\strut{} 25}}%
      \put(814,484){\makebox(0,0){\strut{} 0}}%
      \put(1479,484){\makebox(0,0){\strut{} 1}}%
      \put(2145,484){\makebox(0,0){\strut{} 2}}%
      \put(2810,484){\makebox(0,0){\strut{} 3}}%
      \put(3476,484){\makebox(0,0){\strut{} 4}}%
      \put(4141,484){\makebox(0,0){\strut{} 5}}%
      \put(4807,484){\makebox(0,0){\strut{} 6}}%
      \put(5472,484){\makebox(0,0){\strut{} 7}}%
      \put(6138,484){\makebox(0,0){\strut{} 8}}%
      \put(6803,484){\makebox(0,0){\strut{} 9}}%
      \put(176,2739){\rotatebox{-270}{\makebox(0,0){\strut{}Force $\left[\rm N\right]$}}}%
      \put(4072,154){\makebox(0,0){\strut{}Displacement $\left[\rm mm\right]$}}%
    }%
    \gplgaddtomacro\gplfronttext{%
      \csname LTb\endcsname%
      \put(3718,4514){\makebox(0,0)[r]{\strut{}Standard deviation}}%
      \csname LTb\endcsname%
      \put(3718,4118){\makebox(0,0)[r]{\strut{}100$\%$ r.H. Experiment}}%
      \csname LTb\endcsname%
      \put(3718,3722){\makebox(0,0)[r]{\strut{}100$\%$ r.H. Simulation}}%
    }%
    \gplbacktext
    \put(0,0){\includegraphics{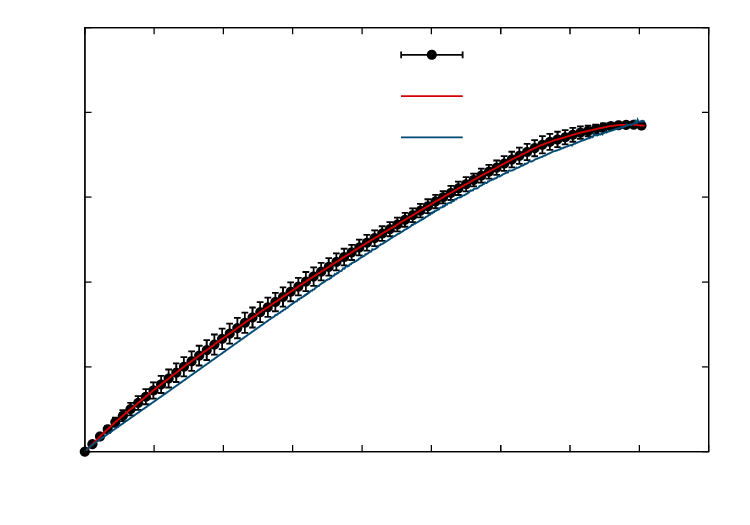}}%
    \gplfronttext
  \end{picture}%
\endgroup

%% file: phasefield/energy.tex
\begingroup
  \makeatletter
  \providecommand\color[2][]{%
    \GenericError{(gnuplot) \space\space\space\@spaces}{%
      Package color not loaded in conjunction with
      terminal option `colourtext'%
    }{See the gnuplot documentation for explanation.%
    }{Either use 'blacktext' in gnuplot or load the package
      color.sty in LaTeX.}%
    \renewcommand\color[2][]{}%
  }%
  \providecommand\includegraphics[2][]{%
    \GenericError{(gnuplot) \space\space\space\@spaces}{%
      Package graphicx or graphics not loaded%
    }{See the gnuplot documentation for explanation.%
    }{The gnuplot epslatex terminal needs graphicx.sty or graphics.sty.}%
    \renewcommand\includegraphics[2][]{}%
  }%
  \providecommand\rotatebox[2]{#2}%
  \@ifundefined{ifGPcolor}{%
    \newif\ifGPcolor
    \GPcolortrue
  }{}%
  \@ifundefined{ifGPblacktext}{%
    \newif\ifGPblacktext
    \GPblacktextfalse
  }{}%
  \let\gplgaddtomacro\g@addto@macro
  \gdef\gplbacktext{}%
  \gdef\gplfronttext{}%
  \makeatother
  \ifGPblacktext
    \def\colorrgb#1{}%
    \def\colorgray#1{}%
  \else
    \ifGPcolor
      \def\colorrgb#1{\color[rgb]{#1}}%
      \def\colorgray#1{\color[gray]{#1}}%
      \expandafter\def\csname LTw\endcsname{\color{white}}%
      \expandafter\def\csname LTb\endcsname{\color{black}}%
      \expandafter\def\csname LTa\endcsname{\color{black}}%
      \expandafter\def\csname LT0\endcsname{\color[rgb]{1,0,0}}%
      \expandafter\def\csname LT1\endcsname{\color[rgb]{0,1,0}}%
      \expandafter\def\csname LT2\endcsname{\color[rgb]{0,0,1}}%
      \expandafter\def\csname LT3\endcsname{\color[rgb]{1,0,1}}%
      \expandafter\def\csname LT4\endcsname{\color[rgb]{0,1,1}}%
      \expandafter\def\csname LT5\endcsname{\color[rgb]{1,1,0}}%
      \expandafter\def\csname LT6\endcsname{\color[rgb]{0,0,0}}%
      \expandafter\def\csname LT7\endcsname{\color[rgb]{1,0.3,0}}%
      \expandafter\def\csname LT8\endcsname{\color[rgb]{0.5,0.5,0.5}}%
    \else
      \def\colorrgb#1{\color{black}}%
      \def\colorgray#1{\color[gray]{#1}}%
      \expandafter\def\csname LTw\endcsname{\color{white}}%
      \expandafter\def\csname LTb\endcsname{\color{black}}%
      \expandafter\def\csname LTa\endcsname{\color{black}}%
      \expandafter\def\csname LT0\endcsname{\color{black}}%
      \expandafter\def\csname LT1\endcsname{\color{black}}%
      \expandafter\def\csname LT2\endcsname{\color{black}}%
      \expandafter\def\csname LT3\endcsname{\color{black}}%
      \expandafter\def\csname LT4\endcsname{\color{black}}%
      \expandafter\def\csname LT5\endcsname{\color{black}}%
      \expandafter\def\csname LT6\endcsname{\color{black}}%
      \expandafter\def\csname LT7\endcsname{\color{black}}%
      \expandafter\def\csname LT8\endcsname{\color{black}}%
    \fi
  \fi
  \setlength{\unitlength}{0.0500bp}%
  \begin{picture}(7200.00,5040.00)%
    \gplgaddtomacro\gplbacktext{%
      \csname LTb\endcsname%
      \put(462,704){\makebox(0,0)[r]{\strut{} 0}}%
      \put(462,1444){\makebox(0,0)[r]{\strut{} 1}}%
      \put(462,2184){\makebox(0,0)[r]{\strut{} 2}}%
      \put(462,2925){\makebox(0,0)[r]{\strut{} 3}}%
      \put(462,3665){\makebox(0,0)[r]{\strut{} 4}}%
      \put(462,4405){\makebox(0,0)[r]{\strut{} 5}}%
      \put(1370,400){\makebox(0,0){\strut{}0\% r.H.}}%
	  \put(2922,400){\makebox(0,0){\strut{}29\% r.H.}}%
	  \put(4475,400){\makebox(0,0){\strut{}67\% r.H.}}%
	  \put(6027,400){\makebox(0,0){\strut{}100\% r.H.}}%
      \put(100,2739){\rotatebox{-270}{\makebox(0,0){\strut{}$E_c$ $\left[\rm N/mm\right]$}}}%
      \put(3830,0){\makebox(0,0){\strut{}Relative humidity $\left[\%\right]$}}%
    }%
    \gplgaddtomacro\gplfronttext{%
    }%
    \gplgaddtomacro\gplbacktext{%
      \csname LTb\endcsname%
      \put(462,704){\makebox(0,0)[r]{\strut{} 0}}%
      \put(462,1444){\makebox(0,0)[r]{\strut{} 1}}%
      \put(462,2184){\makebox(0,0)[r]{\strut{} 2}}%
      \put(462,2925){\makebox(0,0)[r]{\strut{} 3}}%
      \put(462,3665){\makebox(0,0)[r]{\strut{} 4}}%
      \put(462,4405){\makebox(0,0)[r]{\strut{} 5}}%
      \put(1370,400){\makebox(0,0){\strut{}0\% r.H.}}%
      \put(2922,400){\makebox(0,0){\strut{}29\% r.H.}}%
      \put(4475,400){\makebox(0,0){\strut{}67\% r.H.}}%
      \put(6027,400){\makebox(0,0){\strut{}100\% r.H.}}%
      \put(100,2739){\rotatebox{-270}{\makebox(0,0){\strut{}$E_c$ $\left[\rm N/mm\right]$}}}%
      \put(3830,0){\makebox(0,0){\strut{}Relative humidity $\left[\%\right]$}}%
    }%
    \gplgaddtomacro\gplfronttext{%
    }%
    \gplbacktext
    \put(0,0){\includegraphics{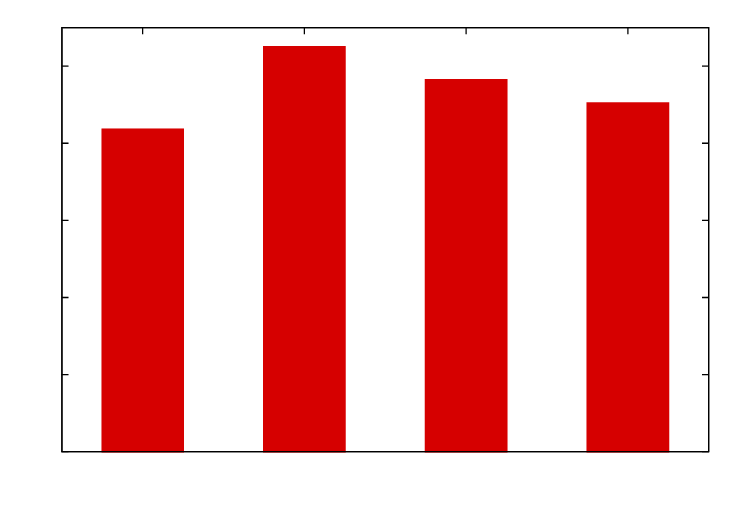}}%
    \gplfronttext
  \end{picture}%
\endgroup